\documentclass[universe,article,accept,pdftex,moreauthors]{Definitions/mdpi} 

\firstpage{1} 
\pubvolume{1}
\issuenum{1}
\articlenumber{0}
\pubyear{2026}
\copyrightyear{2026}
\externaleditor{Firstname Lastname} 
\datereceived{16 June 2026} 
\daterevised{21 July 2026} 
\dateaccepted{23 July 2026} 
\datepublished{ } 

\usepackage{amsmath, amsthm, amssymb} 
\usepackage{epstopdf}
\usepackage{graphicx}
\usepackage{float}
\usepackage{indentfirst}
\usepackage{color}
\usepackage{natbib}
\usepackage[figuresright]{rotating}
\usepackage{enumerate}
\usepackage{appendix}

\providecommand{\dist}{\ensuremath{r}}

\newcommand{\sun}{\hbox{$\odot$}}
\newcommand{\farcs}{.\!\!''}
\newcommand{\farcm}{.\!''}
\newcommand{\arcdeg}{\hbox{$^\circ$}}

\newcommand{\araa}{\textit{Annu. Rev. Astron. Astrophys.}}
\newcommand{\aap}{\textit{Astron. Astrophys.}}
\newcommand{\apj}{\textit{ Astrophys. J.}}
\newcommand{\apjs}{\textit{Astrophys. J. Suppl.}}
\newcommand{\aj}{\textit{ Astron. J.}}
\newcommand{\mnras}{\textit{Mon. Not. R. Astron. Soc.}}
\newcommand{\ao}{\textit{Appl. Opt.}}
\newcommand{\cjaa}{\textit{Chin. J. Astron. Astrophys.}}

\newcommand{\reddash}{\textcolor{black}{--}}

\Title{Resolving Individual Massive Stars in an M31 Star Cluster Exhibiting Wolf--Rayet Features}

\Author{Yunning Zhao $^{1,2}$\orcidA{}, 
Yulong Gao $^{3,4}$,
Wei Zhang $^{1,}$*\orcidB{},
Fuzhen Cheng $^{5}$,
Ming Yang $^{1}$,
Shiming Wen $^{1}$\orcidC{},\linebreak  
Shichao Han $^{1}$\orcidD{}
and Hong Wu $^{1}$
}

 \AuthorNames{Yunning Zhao, Yulong Gao, Wei Zhang, Fuzhen Cheng, Ming Yang, Shiming Wen, Shichao Han and Hong Wu}

\address{%
$^{1}$ \quad National Astronomical Observatories, Chinese Academy of Sciences, Beijing 100101, China; \\
$^{2}$ \quad School of Astronomy and Space Science, University of Chinese Academy of Sciences, Beijing 100049, China\\
$^{3}$ \quad Research Center of Astronomy, Qinghai University, Xining 810016, China; \\
$^{4}$ \quad Department of Physics and Astronomy, Qinghai University, Xining 810016, China\\
$^{5}$ \quad Department of Astronomy, University of  Science and Technology of China, Hefei 230026, China; 
}
\corres{Correspondence: xtwfn@bao.ac.cn}

\abstract{The observed ratio of Wolf--Rayet (WR) star subtypes (e.g., WC/WN) in M31 is significantly higher than the predictions of standard stellar evolution models, yet this key diagnostic discrepancy remains unresolved. This is partly due to incomplete census and the challenge of resolving individual stars in compact clusters with ground-based telescopes. In this work, we report a young cluster with WR features with the (R.A., Dec) = (00:43:39.36, +41:10:08.7), discovered with LAMOST spectra. We use the high spatial resolution of the HST imaging data to resolve the cluster members, leading to the identification of four WR star candidates, four OB stars, and one yellow supergiant. Based on the current known WR sample in M31 observed in the HST F275W filter, about one-third of WRs are contaminated by other sources within a $1\farcs1\times1\farcs1$ aperture, while this fraction can reach up to two-thirds for the LAMOST fiber size of $3\farcs3$. Applying the approach demonstrated in this work to a larger sample of crowded regions in M31 will yield a more complete census and allow for the measurement of a definitive WC/WN ratio, corrected for contamination. Whether this refined ratio increases or decreases the current tension with models remains to be seen, but it will provide the essential observational constraint needed to advance our understanding.}

\keyword{stars: Wolf--Rayet; (stars:) supergiants; (galaxies:) Local Group}

\begin{document}

\section{Introduction}
Wolf--Rayet (WR) stars are the evolved descendants of massive O-type stars characterized by powerful stellar winds (1000--3000 $\rm km\,s^{-1}$) that produce distinctive broad emission line spectra.  
The minimum initial mass required for a star to enter the WR phase is not a fixed value, but~it depends on the adopted stellar-evolution model and physical prescriptions, including mass loss, rotation, internal mixing, binarity, and~metallicity. For~example, in~rotating Geneva models at solar metallicity, rotation can lower the WR formation threshold from about 37 M$_\odot$ to about 22 M$_\odot$ \cite{Meynet-2005}, while enhanced red-supergiant mass loss can allow stars with initial masses around 20 M$_\odot$ to evolve toward WR-like stripped-envelope phases in some models~\cite{vanLoon-2005, Chieffi-2020}. Different theoretical frameworks, such as the Geneva models~\cite{Meynet-2005, Ekstrom-2012}, MESA~\cite{Paxton-2011}, PARSEC~\cite{Bressan-2012}, and~BPASS binary population models~\cite{Eldridge-2008, Eldridge-2016} predict different WR formation channels and lifetimes.
These stars are classified into three main subtypes based on their chemical signatures: WN stars (nitrogen- and helium-dominated, products of CNO-cycle hydrogen burning), WC stars (carbon and oxygen rich, from~triple-$\alpha$ helium burning), and~the rare WO subclass~\cite{Crowther-2007}. Transition-type WN/C stars exhibiting mixed chemistry represent a brief evolutionary phase between WN and WC stages $\sim$1~Myr]~\citep{Crowther-Smith-Willis-1995, Zhang-2020}. Recently, \citet{Sander-2025} proposed that WN stars can evolve directly into the WC stage, identifying the WN/C star as their transitional type.
With extreme mass-loss rates ($\sim$$10^{-5}$--$10^{-4}\,M_{\sun}\, \rm yr^{-1}$) and their eventual fate as core-collapse supernovae or gamma-ray burst progenitors, WR stars serve as crucial laboratories for studying stellar evolution, galactic chemical enrichment, and~feedback mechanisms in star-forming regions~\citep{Callingham-Tuthill-Pope-2019}.

WR features have been detected in extragalactic systems, where they appear as characteristic spectral ``bumps'' due to unresolved emission lines \citep{Allen-1976}. The~``blue bump'' arises from blended broad stellar lines including He II $\lambda$4686, C III/IV $\lambda$4650, and~the N III $\lambda$4634--$\lambda$4641 blend, while the ``red bump'' corresponds to C IV $\lambda$5808-12. Ground-based telescopes cannot resolve individual WR stars in these distant systems, causing their emission features to appear blended in integrated spectra. Such systems are classified as ``WR galaxies'' \citep{Osterbrock-1982}. Their WR populations make them ideal laboratories for studying recent star formation, given WR stars' short lifetimes. Initial studies using SDSS spectra found WR features in only $\sim$0.05\% of galaxies (e.g., \citet{Zhang-2007, Brinchmann-2008, Agienko-2013}), but~this low detection rate reflects SDSS's limitation to central regions or individual HII regions. The~advent of IFU spectroscopy has dramatically improved detection statistics. \citet{Liang-2020} analyzed MaNGA's spatially resolved data for 4621 nearby galaxies, finding WR features in $\sim$2\% of cases---40 times more frequently than SDSS. These datasets also enable investigations of the IMF-metallicity relation, with~evidence suggesting steeper (bottom-heavy) IMFs in higher-metallicity galaxies \citep{Zhang-2007, Liang-2021}.

As one of the nearest star-forming galaxies in the Local Group, M31 has been the subject of extensive studies of its WR population since the pioneering narrow-band imaging surveys of the 1980s \citep{Moffat-1983, Massey-1986}. The~identification of 107 new WR stars, predominantly of the WN type, increased the total number of spectroscopically confirmed WRs in M31 to 154 \citep{Neugent-2012}, a~sample that was further augmented by the first discovery of a WN/C transition star in this galaxy \citep{Shara-2016}. Recent discoveries of 19 additional, primarily reddened but otherwise typical WR stars in limited M31 fields imply that $\sim$60 more may remain undetected, challenging assumptions about the completeness of the current inventory \citep{Neugent-2023}. 

Counting WR stars is essential for testing stellar evolutionary models. The~commonly discussed WR-to-red supergiant (RSG) ratio (e.g., \citet{Massey-2021}) and the WC/WN ratio within WR populations (e.g., \citet{Neugent-2012}) both offer a valuable test for stellar evolutionary models.
{The theoretical predictions of these ratios show a strong dependence on metallicity. Because~metal lines provide an important source of opacity for radiatively driven winds and WR wind mass-loss rates would depend on metallicity, especially on Fe abundance \citep{Vink-2005}. 
At higher metallicity, stronger winds can remove the H-rich and He-rich outer layers more efficiently, affecting both the formation of WR stars and the relative lifetimes of WN and WC phases \citep{Eldridge-2006}. As~a result, WC stars are expected to be more common relative to WN stars in metal-rich environments, while low-metallicity galaxies should have lower WC/WN ratios \citep{Neugent-2019}.}
However, for~the WC/WN ratio, a~persistent discrepancy remains between observed ratios and the predictions of single-star evolutionary models (e.g., the~Geneva models \citep{Meynet-2005, Ekstrom-2012}, MESA models \citep{Paxton-2011}, PARSEC models \citep{Bressan-2012}, {KEPLER models \citep{Woosley-2007} and FRANEC models \citep{Limongi-2018}} ), particularly at high metallicity \citep{Meynet-2005, Neugent-2011}. This disagreement is most pronounced in M31, the~highest-metallicity environment in the Local Group, where the observed ratio is dramatically larger than predicted \citep{Neugent-2019}.~While the use of enlarged WR samples yields a somewhat lower observed WC/WN ratio, it remains significantly higher than the predictions from single-star models at high metallicities \citep{Neugent-2012, Neugent-2023}. Models that incorporate binary evolution, such as the Binary Population and Spectral Synthesis (BPASS) models, provide a better match to the observations \citep{Eldridge-2008, Eldridge-2016}. Yet even these models do not fully resolve the discrepancy, as~their predictions at high metallicity remain lower than the observed values. Furthermore, the~BPASS models do not include the effects of stellar rotation—a factor known to reduce the predicted WC/WN ratio. Accounting for rotation would therefore worsen the existing model-observation discrepancy \citep{Neugent-2019}.

Beyond model limitations, observational biases also affect the detection of the WR population in M31 when relying on ground-based telescopes. First, M31's substantial distance ($\sim$780 kpc) makes its WR stars observationally fainter than those in the Milky Way. This challenge is exacerbated by a known detection bias that favors WC stars over WN stars. Because~the strongest emission line in WCs is nearly four times stronger than that in WNs, making WN stars more difficult to detect at comparable brightness levels \citep{Conti-1989}.  Second, the~crowded environments of dense OB associations complicate the identification of even bright, strongly lined WR stars. In~M33, for~example, about 80\% of WR stars were found within OB associations, with~only 2\% being truly isolated \citep{Neugent-2011}. 
The typical half-light radius of a young cluster is 2--3 pc \cite[][]{Ryon-2015, Brown-2021}, which corresponds to around 2$\farcs$5 at the distance of M31.
Ground-based seeing conditions of M31 (0$\farcs$9 or worse) \citep{Neugent-2023} make it particularly difficult to resolve closely spaced systems where WR stars may have stellar companions or reside in crowded star-forming regions.  Consequently, the~current WR population statistics in M31---including both the total number of identified WR stars and the derived WC/WN ratio---are likely affected by these detection and resolution~limits.

In this paper, we analyze a region exhibiting clear WR features---both blue and red bumps---in the integrated spectra, observed by the Large Sky Area Multi-Object Fiber Spectroscopic
Telescope (LAMOST), combined with photometric data from the Hubble Space Telescope (HST) archive. Our goal is to resolve individual WR stars within the region covered by the fiber of the spectral observation. This method can then be extended to other regions with confirmed WR identifications and available HST observations, enabling a reassessment of the number of WR stars and the WC/WN ratio in~M31.

The paper is structured as follows. In~Section~\ref{sec2},  we describe the observation of the spectrum and the imaging. Data reduction, including the photometry of the cluster, spectral flux calibration, reddening correction, and~the photometry of the star members in the cluster, is presented in Section~\ref{sec3}. The~classification of the massive stars is listed in Section~\ref{sec4}. We give the results and discussion in the last~section.

\section{Observation}
\label{sec2}
\subsection{Integrated Spectrum of the~Cluster}
\label{sec2.1}
LAMOST is a powerful tool for discovering rare objects, thanks to its large field of view (5$\arcdeg$ in diameter) and ability to obtain 4000 spectra in a single exposure \citep{Wang-1996, Su-Cui-2004, Cui-2012, Zhao-2012, Luo-2015}. After~systematically searching spectra in the M31 region, we find a new WR feature spectrum, with~the equatorial coordinates of  (R.A., Dec) = (00:43:39.36, +41:10:08.7), which has been classified as a red supergiant candidate using the strength of the Ca II triplet \citep{Humphreys-1988}, and~has also been classified as a globular cluster \citep{Johnson-2015}. 

\begin{figure}[H]

\includegraphics[width=\textwidth]{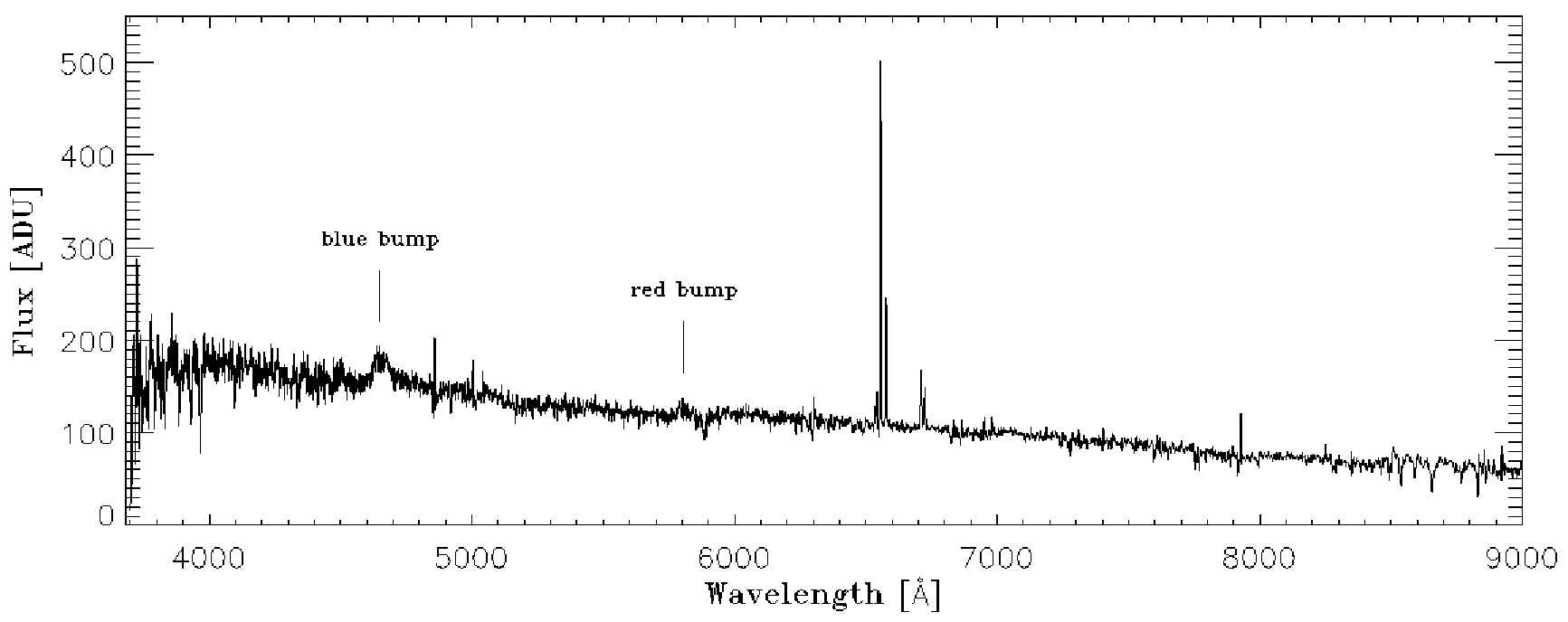}
\caption{Optical spectrum of the target from LAMOST, showing the flux in units of ADU. The~spectrum is characterized by broad WR bumps (blue and red), absorption from an underlying stellar population, and~narrow nebular emission~lines.}
\label{fig:lamost_spectra}
\end{figure}

The spectrum, obtained on 20 November 2011 with an exposure time of 1800s $\times$ 2, is shown in Figure~\ref{fig:lamost_spectra}. The~data were reduced using the standard pipeline, which included bias subtraction, flat-field correction, spectral extraction, wavelength calibration, and~flux calibration \citep{Luo-2015}. This low-resolution spectrum (R $\sim$ 1800) exhibits prominent blue and red bumps centered at approximately 4650 \AA\, and~5808 \AA, respectively.  Due to the limited spectral resolution of LAMOST data, we cannot resolve individual components of these blended features. In~addition to the broad stellar emission lines, the~spectrum shows nebular emission lines ([O II], [O III], H$\alpha$, H$\beta$, [N II] and [S II]) and stellar absorption lines, indicating the presence of ionized gas and other stellar sources with absorption features alongside the WR~stars. 

\begin{figure}[H]

\includegraphics[width=\textwidth]{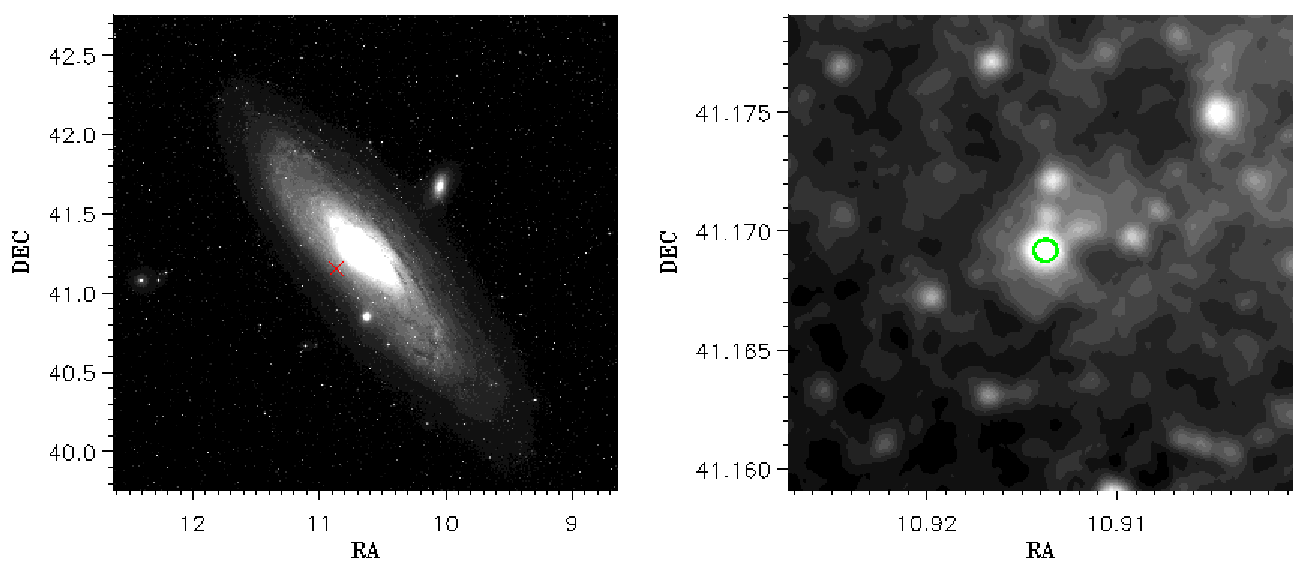}
\caption{DSS2 image of M31. The~left panel has an FOV of $2\arcdeg \times 2\arcdeg$, with~the target cluster marked by a red cross. The~right panel is a zoomed-in view with an FOV of $1\farcm2 \times 1\farcm2$, and~the target is indicated by a green circle with a diameter of  $3\farcs3$.}
\label{fig:simbad_img}
\end{figure}

This cluster is located on the disk of M31, see the red cross in the left panel of Figure~\ref{fig:simbad_img}, with~the image from Digitized Sky Survey 2 (DSS2) with a size of $3\arcdeg\times3\arcdeg$. We zoom in this region to a size of $1\farcm2\times1\farcm2$ in the right panel of Figure~\ref{fig:simbad_img}, and~overplot the position of the LAMOST fiber with a size of $3\farcs3$ in diameter in a green circle. From~this figure, we cannot resolve the members in the fiber and~cannot identify which member contributes to the WR feature shown in the LAMOST~spectrum.

\vspace{-9pt}

\subsection{Hst Photometry~Observation}
\label{sec2.2}
The Panchromatic Hubble Andromeda Treasury (PHAT) surveys obtained imaging of M31, using HST's Wide Field Camera 3 and Advanced Camera for Surveys across six broadband filters: near-ultraviolet (F275W, F336W), optical (F475W and F814W), and~near-infrared (F110W, F160W) \citep{Dalcanton-2012}.  PHAT provides precise astrometry ($<$0$\farcs$1 positional accuracy) and photometry (errors $<$ 0.03 mag for bright sources). The~imaging data were retrieved from the website of The Hubble Legacy Archive (HLA) (\url{https://hla.stsci.edu/}, accessed on 25 July 2026). 
We also obtained three archival F555W images from the same website: one from Proposal ID 10273 and two from Proposal ID 11647. This region was observed between 20 February 2005 and~30 December 2011.~Observations were conducted four times in the F275W, F336W, and~F475W filters, three times in F555W, seven times in F814W, and~twice in F110W and~F160W.

\begin{figure}[H]

\includegraphics[width=\textwidth]{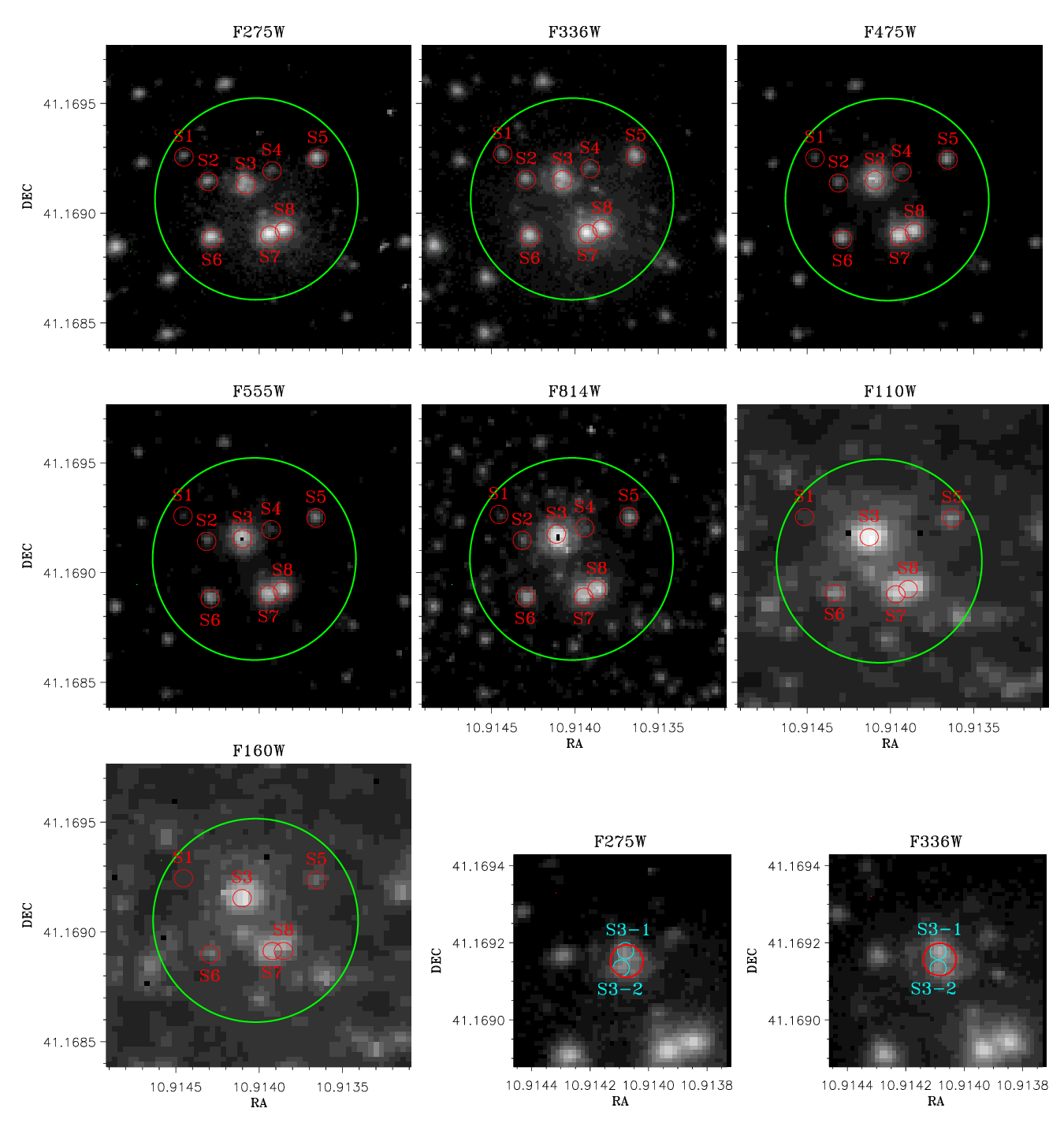}
\caption{Multi-band images from HST observations. The~first seven panels show the field in seven photometric bands, ranging from NUV to NIR. The~green circle, with~a diameter of $3\farcs3$, marks the position of the LAMOST fiber. Sources S1 to S7, detected in these images, are indicated by red apertures with a radius of $0\farcs15$. Source S3, which can be further resolved into S3-1 and S3-2 in the F275W and F336W bands, is identified in the last two panels by cyan circles with a radius of $0\farcs10$.}
\label{fig:HST}
\end{figure}

The seven-band images are shown in Figure~\ref{fig:HST}, with~the green circles marking the positions of the LAMOST fibers ($3\farcs3$ in diameter). Thanks to the high spatial resolution and deep photometry of the HST imaging data, most stellar members within the LAMOST fiber can be resolved and~identified.
\vspace{-3pt}





\section{Data~Analysis}
\label{sec3}

\subsection{Flux~Calibration}
\label{sec3.1}
The LAMOST spectra are relatively flux-calibrated using bright stars (typically F-type stars) from the same plate. However, this method is subject to uncertainty due to the unknown foreground extinction of the calibrating star. To~address this, we performed aperture photometry on HST imaging data to measure magnitudes within the LAMOST fiber aperture. The~results are listed in Table~\ref{tbl:photometry_pars} (see the S$_{\rm tot}$ values).

\begin{figure}[H]

\includegraphics[width=\textwidth]{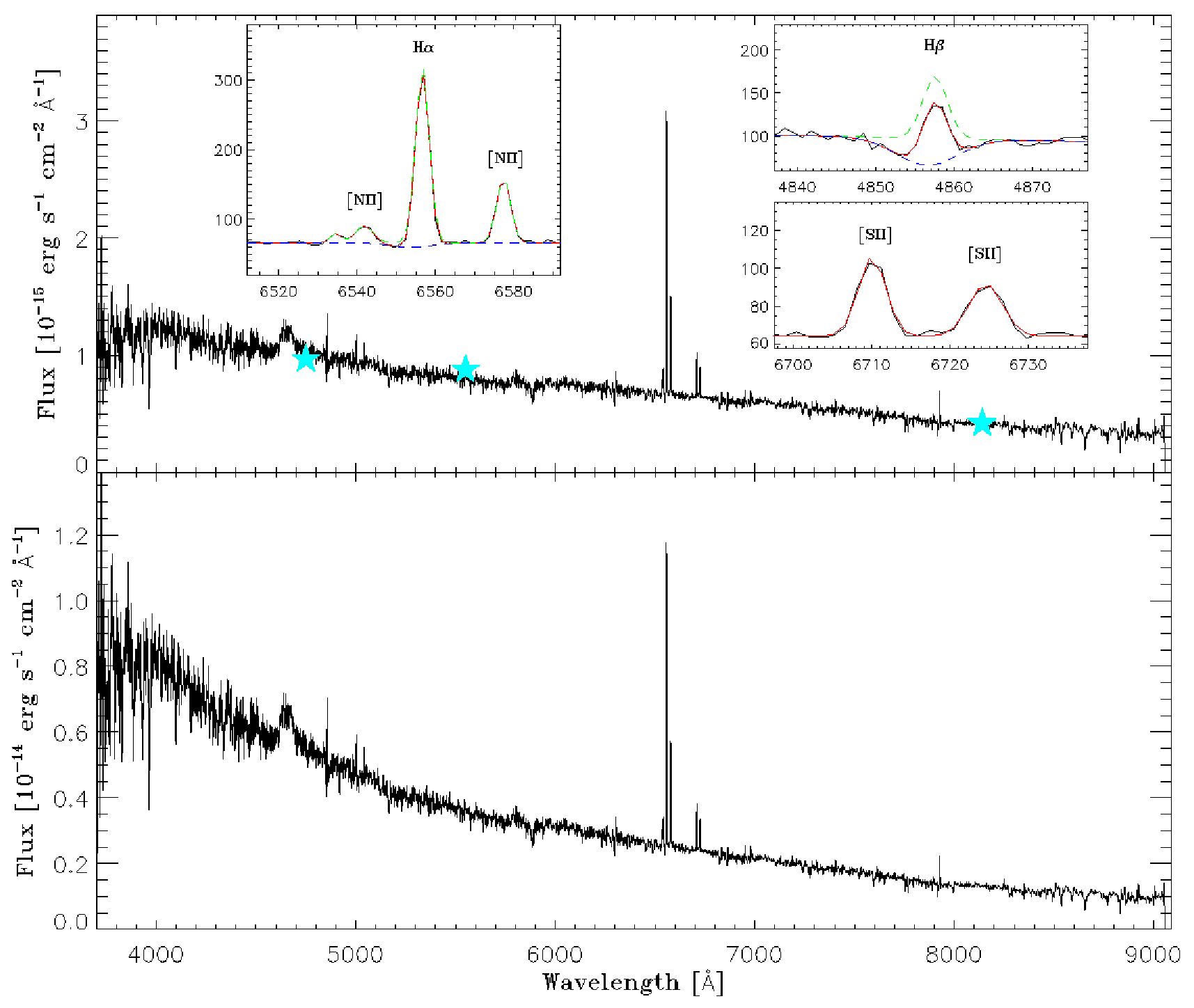}
\caption{(\textbf{Upper}) Re-calibrated spectrum shown in black line over-plotted with the calibrating HST photometry (F475W, F555W, and~F814W bands) indicated by filled cyan stars. The~inserts show multi-Gaussian fits to the [N II], H$\alpha$, H$\beta$, and~[S II] lines, where the nebular, stellar absorption, and~total components are represented by green dashed, blue dashed, and~red solid lines, respectively. (\textbf{Lower}) De-reddened spectrum, derived using the Balmer decrement~method.}
\label{fig:flux_calib_dered}
\end{figure}

For the flux recalibration of the LAMOST spectrum, we selected three HST filters (F475W, F555W, and~F814W) that match its wavelength coverage. The~upper panel of Figure~\ref{fig:flux_calib_dered} shows the recalibrated spectrum, with~the corresponding HST photometric measurements marked as cyan~stars.

\vspace{-9pt}
\begin{table}[H]
\footnotesize
\caption{Photometric data—including source ID and coordinates—for S$_{\rm tot}$, S1--S8, and~the resolved sources S3-1 and S3-2. The~data are listed separately for both the original and de-reddened~values. \label{tbl:photometry_pars}}
\begin{adjustwidth}{-\extralength}{0cm}
\begin{tabularx}{\fulllength}{cCCCCCCCCC}
\toprule
\textbf{(1)} & \textbf{(2)} & \textbf{(3)} & \textbf{(4)} & \textbf{(5)} & \textbf{(6)} & \textbf{(7)} & \textbf{(8)} & \textbf{(9)} & \textbf{(10)} \\
\textbf{OBJ} & \textbf{RA} & \textbf{DEC} &  \textbf{m(F275W)} & \textbf{m(F336W)} & \textbf{m(F475W)} & \textbf{m(F555W)} & \textbf{m(F814W)} & \textbf{m(F110W)} & \textbf{m(F160W)} \\
\midrule
\multicolumn{10}{c}{original} \\
\midrule
S$_{\rm tot}$ &  00 43 39.4 &  41 10 08.7 & 17.48 $\pm$  0.00 & 17.22 $\pm$  0.01 & 16.74 $\pm$  0.02 & 16.51 $\pm$  0.19 & 16.48 $\pm$  0.16 & 16.39 $\pm$  0.01 & 16.68 $\pm$  0.09 \\
\midrule
S1 &  00 43 39.5 &  41 10 09.4 & 23.39 $\pm$  0.03 & 23.31 $\pm$  0.10 & 22.92 $\pm$  0.02 & 23.10 $\pm$  0.02 & 23.37 $\pm$  0.07 & 23.64 $\pm$  0.40 & 24.07 $\pm$  0.36 \\
S2 &  00 43 39.4 &  41 10 08.9 & 22.15 $\pm$  0.02 & 22.08 $\pm$  0.02 & 21.90 $\pm$  0.03 & 21.93 $\pm$  0.06 & 22.10 $\pm$  0.04 & -  & -  \\
S3 &  00 43 39.4 &  41 10 08.9 & 20.52 $\pm$  0.02 & 19.73 $\pm$  0.02 & 18.49 $\pm$  0.01 & 17.66 $\pm$  0.47 & 17.73 $\pm$  0.30 & 17.56 $\pm$  0.06 & 18.01 $\pm$  0.08 \\
S3-1 &  00 43 39.4 &  41 10 09.0 & 21.71 $\pm$  0.11 & 20.38 $\pm$  0.02 & - & - & - & - & -\\
S3-2 &  00 43 39.4 &  41 10 08.8 & 21.68 $\pm$  0.05 & 21.78 $\pm$  0.18 & - & - & - & - & -\\
S4 &  00 43 39.3 &  41 10 09.1 & 23.35 $\pm$  0.02 & 23.11 $\pm$  0.06 & 22.68 $\pm$  0.14 & 22.45 $\pm$  0.10 & 22.51 $\pm$  0.53 & -  & -  \\
S5 &  00 43 39.3 &  41 10 09.3 & 21.29 $\pm$  0.04 & 21.21 $\pm$  0.05 & 20.86 $\pm$  0.03 & 20.88 $\pm$  0.01 & 21.15 $\pm$  0.07 & 21.59 $\pm$  0.06 & 22.52 $\pm$  0.42 \\
S6 &  00 43 39.4 &  41 10 08.0 & 20.13 $\pm$  0.05 & 20.11 $\pm$  0.02 & 20.14 $\pm$  0.03 & 20.20 $\pm$  0.11 & 20.49 $\pm$  0.07 & 21.23 $\pm$  0.05 & 22.02 $\pm$  0.21 \\
S7 &  00 43 39.3 &  41 10 08.1 & 18.69 $\pm$  0.02 & 18.42 $\pm$  0.04 & 18.25 $\pm$  0.05 & 18.30 $\pm$  0.12 & 18.50 $\pm$  0.29 & 19.11 $\pm$  0.21 & 19.64 $\pm$  0.28 \\
S8 &  00 43 39.3 &  41 10 08.1 & 18.97 $\pm$  0.03 & 18.72 $\pm$  0.05 & 18.49 $\pm$  0.04 & 18.52 $\pm$  0.06 & 18.67 $\pm$  0.10 & 19.08 $\pm$  0.19 & 19.70 $\pm$  0.04 \\

\bottomrule
\end{tabularx}
\end{adjustwidth}
\end{table}
\begin{table}[H]\ContinuedFloat
\footnotesize
\caption{\textit{Cont.}}
\begin{adjustwidth}{-\extralength}{0cm}
\begin{tabularx}{\fulllength}{cCCCCCCCCC}
\toprule
\textbf{(1)} & \textbf{(2)} & \textbf{(3)} & \textbf{(4)} & \textbf{(5)} & \textbf{(6)} & \textbf{(7)} & \textbf{(8)} & \textbf{(9)} & \textbf{(10)} \\
\textbf{OBJ} & \textbf{RA} & \textbf{DEC} &  \textbf{m(F275W)} & \textbf{m(F336W)} & \textbf{m(F475W)} & \textbf{m(F555W)} & \textbf{m(F814W)} & \textbf{m(F110W)} & \textbf{m(F160W)} \\

\midrule
\multicolumn{10}{c}{de-reddened} \\
\midrule
S$_{\rm tot}$ &  00 43 39.4 &  41 10 08.7 & 14.79 $\pm$  0.00 & 14.90 $\pm$  0.01 & 14.91 $\pm$  0.02 & 14.83 $\pm$  0.19 & 15.22 $\pm$  0.16 & 15.39 $\pm$  0.01 & 15.87 $\pm$  0.09 \\
\midrule
S1 &  00 43 39.5 &  41 10 09.4 & 20.70 $\pm$  0.03 & 20.99 $\pm$  0.10 & 21.09 $\pm$  0.02 & 21.42 $\pm$  0.02 & 22.10 $\pm$  0.07 & 22.64 $\pm$  0.40 & 23.27 $\pm$  0.36 \\
S2 &  00 43 39.4 &  41 10 08.9 & 19.46 $\pm$  0.02 & 19.76 $\pm$  0.02 & 20.06 $\pm$  0.03 & 20.25 $\pm$  0.06 & 20.83 $\pm$  0.04 & -  & -  \\
S3 &  00 43 39.4 &  41 10 08.9 & 17.83 $\pm$  0.02 & 17.41 $\pm$  0.02 & 16.65 $\pm$  0.01 & 15.98 $\pm$  0.47 & 16.46 $\pm$  0.30 & 16.56 $\pm$  0.06 & 17.21 $\pm$  0.08 \\
S3-1 &  00 43 39.4 &  41 10 09.0 & 19.04 $\pm$  0.11 & 18.06 $\pm$  0.02 & - & - & - & - & -\\
S3-2 &  00 43 39.4 &  41 10 08.8 & 19.01 $\pm$  0.05 & 19.46 $\pm$  0.18 & - & - & - & - & -\\
S4 &  00 43 39.3 &  41 10 09.1 & 20.66 $\pm$  0.02 & 20.79 $\pm$  0.06 & 20.85 $\pm$  0.14 & 20.77 $\pm$  0.10 & 21.24 $\pm$  0.53 & -  & -  \\
S5 &  00 43 39.3 &  41 10 09.3 & 18.60 $\pm$  0.04 & 18.89 $\pm$  0.05 & 19.02 $\pm$  0.03 & 19.20 $\pm$  0.01 & 19.88 $\pm$  0.07 & 20.59 $\pm$  0.06 & 21.72 $\pm$  0.42 \\                                                    
S6 &  00 43 39.4 &  41 10 08.0 & 17.44 $\pm$  0.05 & 17.79 $\pm$  0.02 & 18.30 $\pm$  0.03 & 18.52 $\pm$  0.11 & 19.22 $\pm$  0.07 & 20.23 $\pm$  0.05 & 21.21 $\pm$  0.21 \\
S7 &  00 43 39.3 &  41 10 08.1 & 16.00 $\pm$  0.02 & 16.10 $\pm$  0.04 & 16.41 $\pm$  0.05 & 16.62 $\pm$  0.12 & 17.23 $\pm$  0.29 & 18.11 $\pm$  0.21 & 18.84 $\pm$  0.28 \\
S8 &  00 43 39.3 &  41 10 08.1 & 16.28 $\pm$  0.03 & 16.40 $\pm$  0.05 & 16.65 $\pm$  0.04 & 16.84 $\pm$  0.06 & 17.40 $\pm$  0.10 & 18.08 $\pm$  0.19 & 18.89 $\pm$  0.04 \\
\bottomrule
\end{tabularx}
\end{adjustwidth}
\end{table}

\subsection{Line Measurement and Extinction~Correction}
\label{sec3.2}
The fitting results for H$\alpha$, H$\beta$, [N II] and [S II] lines are displayed in the small insets of the upper panel in  Figure~\ref{fig:flux_calib_dered}, where black lines represent the observed spectrum and red lines show the fitted profiles. For~the [S II] doublet, we used two separate Gaussian components, while the Balmer lines (H$\alpha$ and H$\beta$) required more components due to stellar absorption superposed on nebular emission. Specifically, H$\beta$ was fitted with two Gaussians (nebular emission: cyan dashed line; stellar absorption: blue dashed line), and~H$\alpha$ was treated similarly but included three additional components to account for the [N II] doublet and an unidentified emission feature (centered at $\sim$6534.6 \AA). The~derived radial velocities (RVs), integrated fluxes, and FWHMs are summarized in Table~\ref{tbl:measure_lines}. The~emission-line radial velocities (RVs; $\sim$360 km/s) confirm that the emission nebulae are located in~M31.

\begin{table}[H]

\caption{Measurements of the radial velocity, integrated flux, and~FWHM for the H$\beta$, [N II], H$\alpha$, and~[S II] lines. \label{tbl:measure_lines}}
\begin{tabularx}{\textwidth}{lCCCC}
\toprule       
\textbf{(1)}  & \textbf{(2)}         & \textbf{(3)}        &  \textbf{(4)}  & \textbf{(5)}\\
\textbf{Line} & \textbf{RV}          & \textbf{Flux}       &  \textbf{FWHM}  &  \textbf{Flux} \boldmath{$^d$} \\
     & \textbf{(}\boldmath{$\rm km\,s^{-1}$}\textbf{)}& \textbf{(}\boldmath{$\rm erg\,s^{-1} cm^{-2}$}\textbf{)}  &  \textbf{(\AA)\,} & \textbf{(}\boldmath{$\rm erg\,s^{-1} cm^{-2}$}\textbf{)}\\
\midrule
       H$\beta$ &   $-$317.18 $\pm$      1.57 &    285.86 $\pm$     15.31 &   3.68 $\pm$   0.09  & 1488.12 $\pm$ 79.68 \\
   H$\beta$ $^a$ &   $-$379.09 $\pm$      7.19 &   $-$309.32 $\pm$     15.04 &   9.19 $\pm$   0.34 \\
       {[N II]} &   $-$367.05 $\pm$      4.52 &    140.10 $\pm$      7.98 &   5.23 $\pm$   0.26  & 534.76 $\pm$ 30.47 \\
      H$\alpha$ &   $-$363.99 $\pm$      0.39 &   1117.71 $\pm$      9.58 &   4.14 $\pm$   0.02  & 4257.19 $\pm$ 36.47 \\
  H$\alpha$ $^a$ &   $-$489.83 $\pm$     43.38 &    $-$90.16 $\pm$      0.00 &  12.55 $\pm$   2.21 \\
       {[N II]} &   $-$360.40 $\pm$      0.86 &    412.14 $\pm$      3.98 &   4.18 $\pm$   0.05  & 1565.16 $\pm$ 15.11 \\
       {[S II]} &   $-$368.28 $\pm$      1.86 &    183.42 $\pm$      3.98 &   4.12 $\pm$   0.10  & 683.78 $\pm$ 14.83 \\
       {[S II]} &   $-$362.08 $\pm$      3.13 &    147.84 $\pm$      4.45 &   5.04 $\pm$   0.17  & 550.04 $\pm$ 16.56 \\
\bottomrule
\end{tabularx}
\\
\raggedright
\noindent{\footnotesize{\textsuperscript{a} Stellar absorption component. \textsuperscript{d} De-reddended integrated flux.}}
\end{table}

The H$\beta$ emission in the blue part of the spectrum shows a significant velocity offset compared to the H$\alpha$, [N II], and~[S II] emissions in the red part. This discrepancy likely arises because the wavelength calibration was performed separately for the red and blue spectral regions. Additionally, the~stellar component exhibits a blueshift in RVs relative to the nebular component. However, the~RV values still support their association with M31 rather than being foreground Galactic~stars.  

We measure the observed Balmer decrement $(F_{\mathrm H\alpha}/F_{\mathrm H\beta})_{\rm obs} = 3.91 \pm 0.21$. Using the intrinsic ratio $(F_{\mathrm H\alpha}/F_{\mathrm H\beta})_{\rm int}$ = 2.86 and adopting the extinction law which was introduced by \citet{Charlot-2000}
\begin{equation}
F_{\lambda} = F_{\lambda}^{\mathrm obs} \exp\left[ \tau \left( \frac{\lambda}{5500} \right)^{-0.7} \right]
\end{equation}
The optical depth $\tau$ at 5500 \AA\ ($\tau_{\mathrm V}$) is derived from
\begin{equation}
\tau_{\mathrm V} = - \frac{\ln(F_{\mathrm H\alpha}/F_{\mathrm H\beta})_{\mathrm obs} - \ln(F_{\mathrm H\alpha}/F_{\mathrm H\beta})_{\rm int}}{\mathrm (\lambda_{H\alpha}/5500)^{-0.7} - (\lambda_{H\beta}/5500)^{-0.7}}
\end{equation}
This yields $\tau_{\rm V} = 1.51 \pm 0.27$.

The attenuation derived from this method corresponds to that of the gas, i.e.,~ $E(B-V)_{\rm{gas}} = 2.5\,\tau_V$. The~stellar attenuation            
$E(B-V)_{\rm{star}}$, however, is not always identical to it. The~ratio $f = E(B-V)_{\rm{star}} / E(B-V)_{\rm{gas}}$ typically ranges from about    
0.44 to $\sim$1. \mbox{\citet{Li-2021}} further reported an anti-correlation between $E(B-V)_{\rm{star}}$ and the luminosity-weighted stellar age $t_L$.     
Because the region studied in this work is young, the~value of $f$ should be significantly larger than 0.44. For~simplicity, we assume $f = 1$ in this region and
apply the corresponding extinction correction to the observed spectrum, broadband magnitudes, and~emission-line fluxes.
The bottom panel of Figure~\ref{fig:flux_calib_dered} presents the de-reddened spectrum. The~de-reddened total magnitudes for each band are summarized in Table~\ref{tbl:photometry_pars}, while the de-reddened emission-line fluxes are listed in column 5 of Table~\ref{tbl:measure_lines}.

\subsection{Property of the Ionized~Medium}
\label{sec3.3}
Using the de-reddened [S II] $\lambda6717$/$\lambda6731$ line ratio of $1.24 \pm 0.06$ and adopting electron temperature $T_{\rm e}$ = 10,000~K, we applied Equation~(3) of \citet{Proxauf-2014} to derive an electron density of $n_{\rm e} = 136 \pm 56$~cm$^{-3}$. We also estimated the oxygen abundance using two independent methods: the N2 method \citep{PP04}, which yielded 12~+~log(O/H)~=~$8.652~\pm~0.003$, and~the N2S2H$\alpha$ method \citep{Dopita-2016}, which gave 12~+~log(O/H)~=~$8.76 \pm 0.01$. Adopting the solar oxygen abundance 12 + log(O/H) = 8.69 from \citet{Asplund-2009}, these values indicate that this region has an approximately solar oxygen~abundance.

\subsection{Photometry of HST Images for Resolved~Targets}
\label{sec3.4}
Thanks to HST's superior spatial resolution, we performed photometry on resolved targets within 3$\farcs$3  regions. Using an fixes aperture radius of 0$\farcs$15  (red circles in Figure~\ref{fig:HST}), we identified and measured eight targets (labeled S1--S8). Additionally, we detected numerous faint sources in F814W that were absent in shorter wavelength filters, indicating red spectral energy distributions. These objects were undetectable in longer wavelength filters (F110W and F160W) due to their lower spatial resolution. As~our study focuses on massive stars, we excluded these faint, red objects from subsequent analysis. The~exceptional resolution of the F275W and F336W filters allowed us to resolve S3 into two distinct components (S3-1 and S3-2, marked by cyan circles in the last two panels of Figure~\ref{fig:HST}), for~which we used a smaller 0$\farcs$075 aperture for precise photometry. The~measured magnitudes are presented in Table~\ref{tbl:photometry_pars}, along with de-reddened magnitudes calculated using the extinction correction method described in Section~\ref{sec3.2}. The~uncertainties of magnitudes are derived from the scatter of multi-epoch observation in each~band.

\begin{figure}[H]

\includegraphics[width=\textwidth]{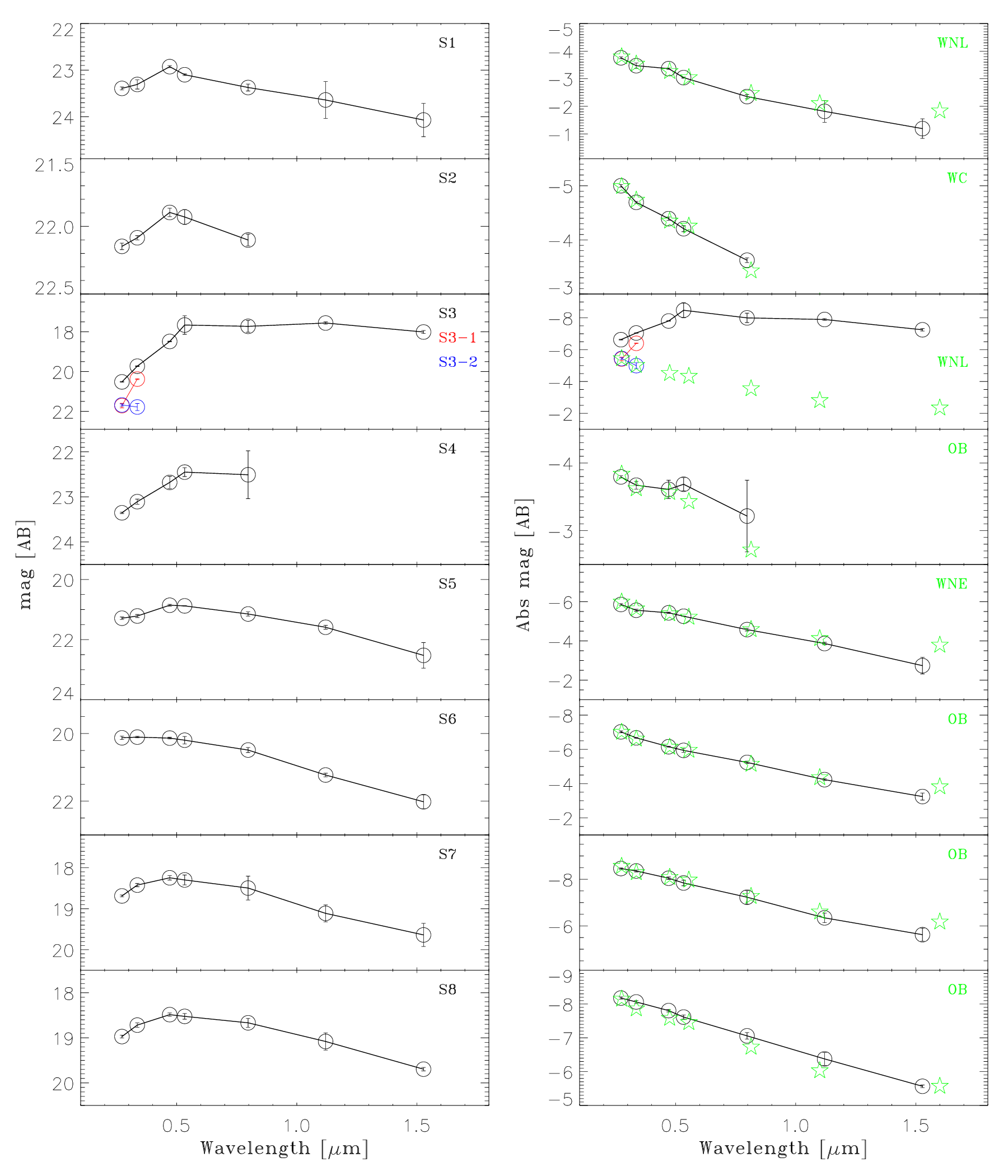}
\caption{(\textbf{Left}) Observed SEDs (black circles with lines) for the sources, with~the resolved components S3-1 and S3-2 colored red and blue. (\textbf{Right}) De-reddened SEDs in absolute magnitudes, compared with the best-fitting PoWR models (open green stars). Each panel is annotated with the source ID or classification in the upper-right~corner.}
\label{fig:SED}
\end{figure}

Figure~\ref{fig:SED} displays the spectral energy distributions (SEDs) for the nine objects. The~left panels present observed magnitudes (open black circles), while the right panels show absolute magnitudes (open black circles) corrected for extinction and adjusted to a distance modulus of 24.46 for M31. Sources S1, S5--S8 are detected in all seven bands, whereas S2 and S4 lack detections in F110W and F160W. S3 appears in all bands (open black circles), while its resolved components S3-1 and S3-2 (red and blue open circles, respectively) are only detected in the higher-resolution F275W and F336W~bands.
\vspace{-6pt}

\section{Identifying the Spectra~Types}
\label{sec4}
Although we lack individual spectra for each star, by~combining multi-band photometry (from FUV to NIR) with the integrated LAMOST spectrum, we can still constrain spectral types and identify potential WR stars contributing to the observed WR features (both red and blue bumps).

Except S3, the~SEDs of all targets display characteristically blue slopes, consistent with early-type stellar photospheres. We determine fundamental stellar parameters by fitting these SEDs with synthetic spectra from the Potsdam Wolf--Rayet (PoWR) model atmospheres \citep{Grafener-2002, Hamann-2003, Sander-2015}, which incorporate non-local thermal equilibrium (non-LTE) physics, spherical expansion, and~metal line blanketing. For~this study, we utilize the publicly available model grids with fixed Galactic metallicity from the PoWR database (\url{http://www.astro.physik.uni-potsdam.de/PoWR.html}, accessed on 25 July 2026), including: (1) WN star models \citep{Hamann-Grafener-2004,Todt-2015}, (2) WC star models \citep{Sander-Hamann-Todt-2012}, and~(3) OB star models \citep{Hainich-2019}. This metallicity is justified by the oxygen abundance measured in Section~\ref{sec3.3}, which is comparable to the solar value.
We compare the observed SEDs to the model grids and select the model with the minimum $\chi^2$ as the best fit.
The best-fitting results are shown as green stars in the right panels of Figure~\ref{fig:SED}, with~the corresponding classifications indicated in the upper right corners.
We also list their classifications and fitting stellar parameters in Table~\ref{tbl:fitting_pars}. 
The uncertainties of the fitted stellar parameters are estimated by propagating the photometric uncertainties, which are derived from the scatter of the multi-epoch HST measurements in each band, as~described in Sections~\ref{sec2.2} and \ref{sec3.4}. The~masses listed in Table~\ref{tbl:fitting_pars} are current stellar masses inferred from the best-fitting PoWR atmosphere models, rather than initial or zero-age main-sequence (ZAMS) masses. We note that the derived parameters are model-dependent, because~they are based on broadband SED fitting and the assumptions of the adopted PoWR atmosphere-model grids. In~particular, the~mass-loss rates and terminal velocities are constrained by the selected PoWR models rather than directly measured from spatially resolved stellar spectra.
Our SED fitting results indicate that sources S1, S2, and~S5 are WR stars candidates, while S4, S6, S7, and~S8 are OB stars. It should be noted that these WR star classifications are based solely on broadband SED fitting, so we regard them as WR star candidates. Definitive certification of these WR sources require more spatially resolved spectroscopic observations in the~future.

For S3, we resolve it into S3-1 and S3-2 using imaging data in previous section. S3-2 exhibits blue slope, while S3-1 shows red slope.
Although the object S3-2 has only two bands of observational data, we still perform SED fitting as other early-type sources. We also present the fitting results across all seven bands, as~we will subtract its contribution from the SED of S3 to isolate the SED of S3-1.
For S3-2, the~classification is ambiguous due to having only two UV photometric bands. To~resolve this, we combine the modeled WR spectra of S1, S2, and~S5 to compute the expected blue and red bump fluxes. The~results indicated a significant deficit, requiring an additional WR contribution. Consequently, we classify S3-2 as a WR star candidate, with~a template fit identifying it as a WNL subtype. The~classification and stellar parameters of S3-2 are also listed in Table~\ref{tbl:fitting_pars}.
Including S3-2 with the other three WR stars yields integrated blue and red bump fluxes of $1.9~\times~10^4~\rm{erg~s^{-1}~cm^{-2}}$ and $3.6 \times 10^5~\rm{erg~s^{-1}~cm^{-2}}$, respectively, which are consistent with the values of $(4.2\, \pm 0.4) \times 10^4~\rm{erg~s^{-1}~cm^{-2}}$ and $(2.6\, \pm 2.5) \times 10^5~\rm{erg~s^{-1}~cm^{-2}}$ derived from the observed LAMOST~spectra.

After subtracting the contribution of S3-2 from S3, the~resulting SED for S3-1 exhibits a red shape. Its absolute magnitude ($\rm M(F555W) = -8.47 \pm 0.47$) suggests a supergiant. Initial attempts to fit the entire SED with standard templates were unsuccessful, likely due to excess UV emission from an unresolved companion—a common feature in binary systems containing supergiants \citep{Dai-2025}. Therefore, we uSED the VO SED Analyzer (VOSA)~\citep{Bayo-2008} to fit only the longer-wavelength portion of the SED (green line in Figure~\ref{fig:S3-1-fit}) with ATLAS9  models \citep{Castelli-2003}. A~good match was obtained with a yellow supergiant template characterized by $T_{\rm eff} = 6500 \pm 125$ K, $\log g = 0.50 \pm 0.40$, and~$\log(L/L_\odot) = 5.050 \pm 0.002$.~As the most luminous cluster member, S3-1 dominates the red portion of the integrated light. Although~its UV classification remains uncertain due to limited data, this has a negligible impact on the longer-wavelength analysis.
Our analysis thus reveals that the cluster's prior classification as a red supergiant stems from the overwhelming dominance of this single supergiant's~light.

\textls[-15]{We checked the yellow and red supergiant catalogs of \citet{Drout-2009}, \mbox{\citet{Gordon-2016}},} and~\citet{Wu-2025}, and~found that S3-1 is not included in these catalogs. Since S3-1 is located within the same compact stellar system, we consider it likely to be physically associated with the system, although~a chance projection along the line of sight cannot be completely ruled out without spatially resolved spectroscopy or kinematic information. If~S3-1 is a coeval member, its coexistence with the WR/OB candidates may be explained by different evolutionary timescales for stars with different initial~masses.

As we only identify the brightest stars in the LAMOST fiber, and~there are also contributions from fainter stars and ionized gas within the same fiber, we did not attempt to combine the SEDs of all individual stars for comparison with the observed LAMOST spectrum.
We must acknowledge that the current SED composition may not be unique, as~certain subtypes of WR and OB stars often exhibit similar SEDs due to their comparable luminosities and temperatures. We consider this the most probable composition, and~we look forward to higher-spatial-resolution spectra to confirm the stars within this~cluster.

\begin{table}[H]

\fontsize{9}{9}\selectfont
\caption{Stellar classification and physical parameters for the observed sources (S1, S2, S4--S8) and the resolved sub-component~S3-2. \label{tbl:fitting_pars}}
\begin{tabularx}{\textwidth}{cccccccc}
\toprule
\textbf{(1)} & \textbf{(2)}  & \textbf{(3)}  &  \textbf{(4)}      & \textbf{(5)}             & \textbf{(6)}        &   \textbf{(7)}               &   \textbf{(8)}        \\
\textbf{OBJ} & \textbf{Type} & \textbf{T}\boldmath{$_*$} & \textbf{log R}\boldmath{$_{\rm t}$} & \textbf{log L}           & \textbf{M}\boldmath{$_*^1$}         & \textbf{log \.M}            & \boldmath{$v_\infty$} \\
    &      &  \textbf{(kK)}  & \textbf{(R}\boldmath{$_\odot$)} & \textbf{(L}\boldmath{$_\odot$}\textbf{)}       & \textbf{(M}\boldmath{$_\odot$}\textbf{)}  &\textbf{ (M}\boldmath{$_\odot$ }\textbf{yr}\boldmath{$^{-1}$}\textbf{)}   & \textbf{(km s}\boldmath{$^{-1}$}\textbf{)}\\                 
\midrule
  S1   &    WNL &  99 $\pm$  1        &  0.70 $\pm$  0.02 &  5.30 $\pm$  \reddash  & 29.90 $\pm$  \reddash & $-$5.49 $\pm$  0.02     &   1000 $\pm$     \reddash \\
  S2   &    WC  &  70 $\pm$  1        &  0.81 $\pm$  0.03 &  5.30 $\pm$  \reddash  & 11.60 $\pm$  \reddash & $-$5.09 $\pm$  0.02     &   2000 $\pm$     \reddash \\
  S3-2 &    WNL &  50 $\pm$  \reddash &  1.55 $\pm$  0.11 &  5.30 $\pm$  \reddash  & 29.90 $\pm$  \reddash & $-$5.86 $\pm$  0.17     &    999 $\pm$     \reddash \\
  S4   &    OB  &  19 $\pm$  \reddash &  2.26 $\pm$  0.02 &  3.96 $\pm$  \reddash  &  9.19 $\pm$  \reddash & $-$7.00 $\pm$  0.00     &    863 $\pm$    91        \\
  S5   &    WNE &  46 $\pm$  2        &  0.84 $\pm$  0.11 &  5.30 $\pm$  \reddash  & 12.00 $\pm$  \reddash & $-$4.48 $\pm$  0.08     &   1600 $\pm$     \reddash \\
  S6   &    OB  &  43 $\pm$  3        &  2.87 $\pm$  0.02 &  5.80 $\pm$  0.10      & 59.54 $\pm$ 12.05     & $-$7.00 $\pm$  \reddash &   3039 $\pm$   358        \\
  S7   &    OB  &  22 $\pm$  2        &  3.18 $\pm$  0.06 &  5.82 $\pm$  0.08      & 46.44 $\pm$  5.90     & $-$7.00 $\pm$  \reddash &   1212 $\pm$   351        \\
  S8   &    OB  &  27 $\pm$  2        &  3.13 $\pm$  0.01 &  5.89 $\pm$  0.10      & 52.57 $\pm$  8.24     & $-$7.00 $\pm$  \reddash &   1668 $\pm$   161        \\
\bottomrule         
\end{tabularx}
\\
\raggedright
\noindent{\footnotesize{\textsuperscript{1} Current stellar mass.}}\\
\end{table}
\vspace{-15pt}

\begin{figure}[H]

\includegraphics[width=\textwidth]{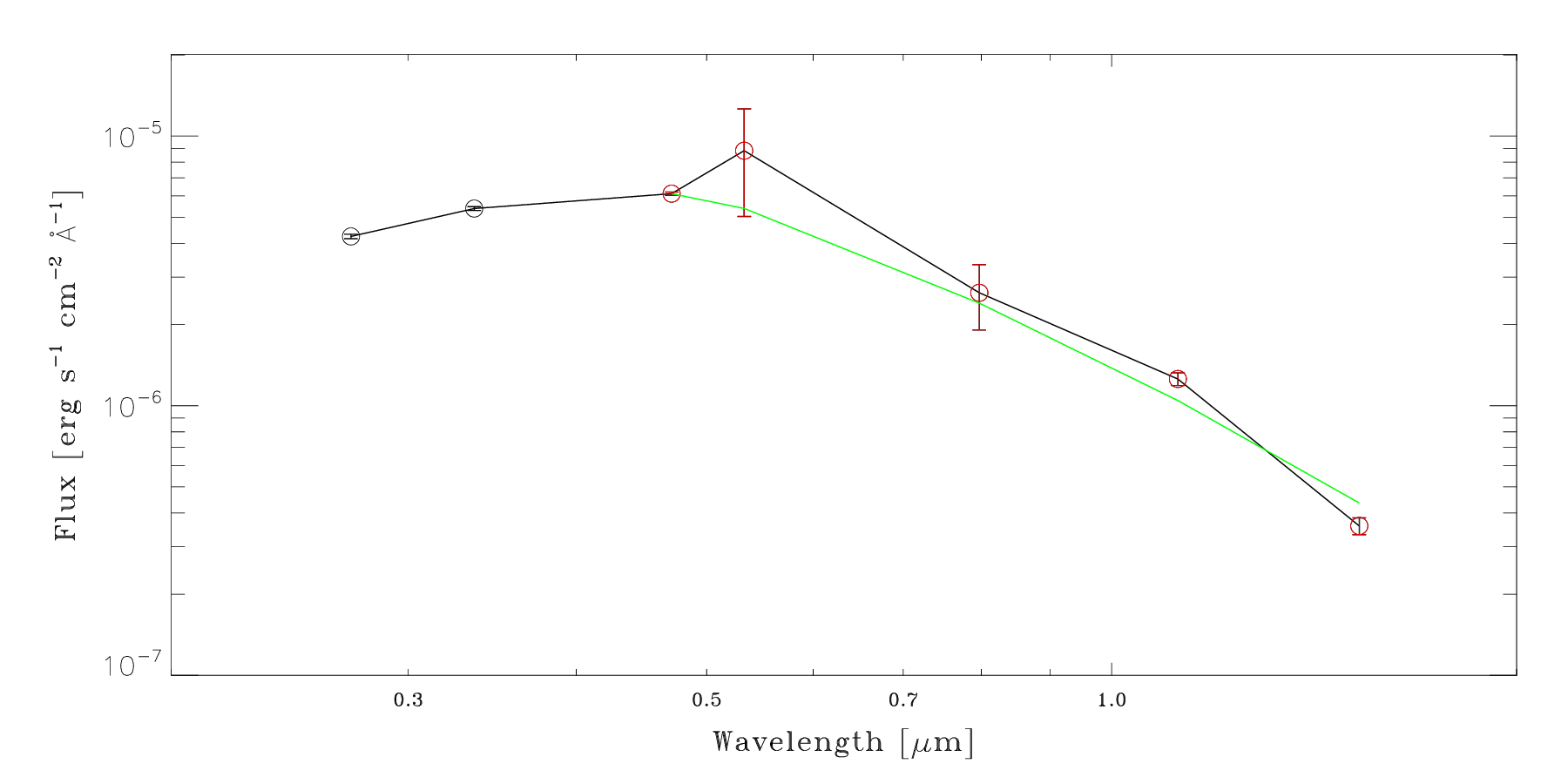}
\caption{SED of S3-1. The~observed data (open circles with black line) is fitted by a model (green line) using the data points highlighted in~red.}
\label{fig:S3-1-fit}
\end{figure}

\section{Summary and~Discussion}

\begin{itemize}

\item We have discovered a cluster in M31 that exhibits WR features using LAMOST spectra. While previous ground-based observations identified this cluster as a single supergiant star based on its SED, our analysis of high spatial resolution HST multi-band data (spanning UV to IR) resolves nine individual members: four WR star candidates, four OB stars, and~one yellow supergiant.  The~presence of these massive stars suggests that the region is young and has recently undergone a new process of star~formation.

\item The yellow supergiant dominates the cluster's emission at red wavelengths, explaining why earlier ground-based studies misclassified the entire cluster as a single supergiant. Furthermore, our inability to fully fit the UV emission from this supergiant suggests the presence of an unresolved~companion.

\item These results highlight the crucial role of high spatial resolution in properly identifying cluster members in M31. They also imply that binary fraction estimates for supergiants in M31 based on ground-based data are affected by such unresolved~systems.


\end{itemize}

To better determine the WC/WN ratio in nearby galaxies like M31, M33, LMC, and~SMC, and~to investigate its metallicity dependence while comparing with theoretical models, it is necessary to re-examine regions showing WR features in ground-based spectra and to identify unresolved WR stars using existing HST or JWST data. These preparatory studies will establish a reliable census of WR populations for detailed~analysis.

The Habitable Worlds Observatory (HWO) will revolutionize massive star studies by enabling panchromatic, spatially resolved ($\leq$0$\farcs$015) observations of individual stars across the Local Volume ($\sim$5 Mpc) \citep{Smercina-2025}. With~resolution surpassing HST and JWST, HWO will provide unprecedented precision in characterizing WR populations, allowing us to accurately measure WC/WN ratios across metallicities, directly test stellar evolution models, and~reveal how massive star evolution varies in different galactic environments. This capability will open a new era in understanding WR stars and their role in galactic~ecosystems.

\vspace{6pt}
\authorcontributions{Conceptualization and methodology, W.Z.; software, W.Z. and  Y.Z.; validation, Y.G. and M.Y.; formal analysis, Y.Z. and W.Z.; investigation, Y.Z., W.Z. and S.W.; data curation, W.Z., Y.G. and M.Y.; writing---original draft preparation, Y.Z. and W.Z.; writing---review and editing, Y.Z., W.Z., Y.G., M.Y. and S.H.; visualization, Y.Z.; supervision, F.-Z.C. and H.W.; project administration, W.Z. and H.W.; funding acquisition, W.Z. and H.W. All authors have read and agreed to the published version of the~manuscript.}
\funding{This work is supported by the National Key R\&D Program of China grants (Nos. 2021YFA1600401 and 2021YFA1600400) and the National Natural Science Foundation of China (NSFC) (Nos. 12090041, and~12090040). This work is also sponsored by the Strategic Priority Research Program of the Chinese Academy of Sciences (No. XDB0550100).}

\dataavailability{This work made use of the data from LAMOST (Large Sky Area Multi-Object Fiber Spectroscopic Telescope, also known as the Guoshoujing Telescope) (\url{https://cstr.cn/31118.02.LAMOST}, accessed on 25 July 2026). LAMOST is a Chinese national mega-science facility, operated by the National Astronomical Observatories, Chinese Academy of Sciences. 
This publication makes use of VOSA, developed under the Spanish Virtual Observatory (\url{https://svo.cab.inta-csic.es}, accessed on 25 July 2026) project funded by MCIN/AEI/10.13039/501100011033/ through grant PID2020-112949GB-I00. VOSA has been partially updated by using funding from the European Union's Horizon 2020 Research and Innovation Programme, under~grant agreement No. 776403 (EXOPLANETS-A).}

\acknowledgments{We thank the anonymous reviewers for their constructive comments and suggestions, which helped improve the clarity and quality of this manuscript. W.Z. thanks Cheng Li and Yang Huang for helpful~suggestions.}
\conflictsofinterest{The authors declare no conflicts of~interest.}
%


\appendixtitles{no} 
\appendixstart
\appendix
\section[\appendixname~\thesection]{}
To systematically determine what fraction of WR stars in M31 might be contaminated when observed with a LAMOST-sized fiber, we used a combined WR catalog from \mbox{\citet{Neugent-2012}}, \citet{Shara-2016}, and~\citet{Neugent-2023}, containing a total of 174 unique WR~stars.

We cross-matched this sample with the HST archive data and found that 91 of these stars have been observed with the F275W filter. 
In this section, we only count the number of bright stars in the images, and~do not perform SED fitting,  as~most of these WR stars are either not observed by LAMOST or do not have high signal-to-noise spectra. By~visually inspecting these images in Figure~\ref{fig:merged_hstimg1}, we found that 34 out of 91 (37.4\%) appear isolated within a diameter circular aperture, meaning no other bright sources are present. The~remaining 57 sources (62.6\%) show two or more bright objects within the same aperture, implying that spectroscopic observations with LAMOST would suffer from significant contamination from nearby stars. This will potentially affect the WR stars census and the analysis of the WC/WN ratio in~M31.

We repeated this analysis using a smaller $1\farcs1\times1\farcs1$ square aperture to simulate a higher-resolution fiber. Within~this smaller region, 63 WR stars (69.2\%) appear single, while 28 (30.8\%) still show evidence of~contamination.

The catalog of unique WR stars in M31 used in this analysis is presented in Table~\ref{tbl:obs_infor}. For~each source, the~table provides an ID (1–174), J2000 coordinates (RA and Dec), and~a flag (Y/N) denoting availability of HST F275W imaging. Columns 5 and 6 give the results of our visual contamination assessment for two aperture sizes: a LAMOST-sized fiber and a smaller $1\farcs1\times1\farcs1$ square aperture, with~entries labeled `single' or `multiple'.

\begin{table}[H]

\setlength{\tabcolsep}{4pt}
\caption{Unique sample of WR stars in M31. The~columns are as follows: (1) the ID number (ranging from 1 to 174); (2) and (3) the J2000 RA and Dec; (4) a flag (Y/N) indicating whether the star has been observed with the HST F275W filter; (5) a classification flag (S/M: single/multiple) based on a visual inspection for contamination within a LAMOST-sized fiber aperture; (6) the same classification as column 5, but~for a $1\farcs1\times1\farcs1$ square aperture; and (7) the reference catalog: N12,~\cite{Neugent-2012}; N23,~\cite{Neugent-2023}; S16,~\cite{Shara-2016}.
\label{tbl:obs_infor}}
\begin{tabularx}{\textwidth}{CCCCCCC}
\toprule
\textbf{(1)}          & \textbf{(2)}     & \textbf{(3)}   & \textbf{(4)}         & \textbf{(5)}           & \textbf{(6)}            & \textbf{(7)} \\
\textbf{ID}           & \textbf{RA}      & \textbf{Dec}   & \textbf{F275W}       & \textbf{n(}\boldmath{$3\farcs3$}\textbf{)} & \textbf{n(}\boldmath{$1\farcs1$}\textbf{)}  & \textbf{Ref} \\
\midrule
001          & 00:38:57.19 & +40:31:32.2 & N     & -  & -  & N12 \\
002          & 00:39:11.04 & +40:38:17.5 & N     & -  & -  & N12 \\
003          & 00:39:19.53 & +40:22:11.6 & N     & -  & -  & N12 \\
004          & 00:39:33.46 & +40:20:18.9 & Y     & S  & S  & N12 \\
005          & 00:39:35.63 & +40:28:11.5 & N     & -  & -  & N12 \\
006          & 00:39:39.61 & +40:20:31.1 & Y     & M  & S  & N12 \\
007          & 00:39:39.97 & +40:34:50.4 & N     & -  & -  & N12 \\
008          & 00:39:45.82 & +40:23:03.2 & N     & -  & -  & N12 \\
009          & 00:39:48.84 & +40:52:56.3 & N     & -  & -  & N12 \\
010          & 00:40:02.84 & +40:38:52.8 & N     & -  & -  & N12 \\
011          & 00:40:05.65 & +40:58:48.1 & N     & -  & -  & N12 \\
012          & 00:40:19.47 & +40:52:24.9 & N     & -  & -  & N12 \\
013          & 00:40:20.44 & +40:48:07.7 & N     & -  & -  & N12 \\
014          & 00:40:21.13 & +40:35:20.6 & N     & -  & -  & N12 \\
015          & 00:40:22.43 & +40:52:34.6 & N     & -  & -  & N12 \\
016          & 00:40:23.02 & +40:44:54.1 & Y     & S  & S  & N12 \\
017          & 00:40:24.33 & +40:50:16.2 & N     & -  & -  & N12 \\
018          & 00:40:26.23 & +40:44:59.6 & Y     & M  & M  & N12 \\
019          & 00:40:29.27 & +40:39:16.6 & N     & -  & -  & N12 \\
020          & 00:40:31.67 & +40:39:09.0 & N     & -  & -  & N12 \\
021          & 00:40:32.58 & +40:35:50.1 & N     & -  & -  & N12 \\
022          & 00:40:34.17 & +40:43:40.4 & N     & -  & -  & N12 \\
023          & 00:40:34.69 & +40:44:32.9 & Y     & S  & S  & N12 \\
024          & 00:40:35.18 & +40:38:38.5 & N     & -  & -  & N12 \\
025          & 00:40:36.76 & +41:01:04.3 & N     & -  & -  & N12 \\
026          & 00:40:43.28 & +40:35:25.2 & N     & -  & -  & N12 \\
027          & 00:40:56.49 & +41:03:08.7 & Y     & S  & S  & N12 \\
028          & 00:40:56.72 & +41:02:55.4 & Y     & S  & S  & N12 \\
029          & 00:40:58.49 & +41:04:14.8 & Y     & S  & S  & N12 \\
030          & 00:40:59.13 & +40:36:52.0 & N     & -  & -  & N12 \\

\bottomrule
\end{tabularx}
\end{table}
\begin{table}[H]\ContinuedFloat
\caption{\textit{Cont.}}
\begin{tabularx}{\textwidth}{CCCCCCC}
\toprule
\textbf{(1)}          & \textbf{(2)}     & \textbf{(3)}   & \textbf{(4)}         & \textbf{(5)}           & \textbf{(6)}            & \textbf{(7)} \\
\textbf{ID}           & \textbf{RA}      & \textbf{Dec}   & \textbf{F275W}       & \textbf{n(}\boldmath{$3\farcs3$}\textbf{)} & \textbf{n(}\boldmath{$1\farcs1$}\textbf{)}  & \textbf{Ref} \\
\midrule

031          & 00:41:00.97 & +40:36:33.0 & N     & -  & -  & N12 \\
032          & 00:41:02.04 & +41:04:46.0 & N     & -  & -  & N12 \\
033          & 00:41:07.31 & +41:04:17.0 & N     & -  & -  & N12 \\
034          & 00:41:09.46 & +40:49:07.8 & Y     & M  & M  & N12 \\
035          & 00:41:09.59 & +40:48:56.2 & Y     & M  & S  & N12 \\
036          & 00:41:14.95 & +40:44:48.2 & N     & -  & -  & N12 \\
037          & 00:41:15.95 & +41:09:06.8 & N     & -  & -  & N12 \\
038          & 00:41:26.11 & +41:12:20.0 & Y     & M  & S  & N12 \\
039          & 00:41:26.39 & +41:12:03.5 & Y     & M  & M  & N12 \\
040          & 00:41:29.11 & +41:13:11.4 & Y     & M  & M  & N12 \\
041          & 00:41:29.17 & +41:13:35.0 & Y     & M  & M  & N12 \\
042          & 00:41:30.37 & +41:05:00.9 & N     & -  & -  & N12 \\
043          & 00:41:34.99 & +41:05:52.3 & N     & -  & -  & N12 \\
044          & 00:41:41.81 & +40:37:11.5 & N     & -  & -  & N12 \\
045          & 00:41:43.00 & +41:18:59.7 & N     & -  & -  & N12 \\
046          & 00:41:43.09 & +40:40:45.9 & N     & -  & -  & N12 \\
047          & 00:41:44.47 & +40:45:17.1 & N     & -  & -  & N12 \\
048          & 00:41:45.33 & +41:21:04.0 & N     & -  & -  & N12 \\
049          & 00:41:47.24 & +41:06:47.5 & N     & -  & -  & N12 \\
050          & 00:41:48.27 & +41:17:39.9 & N     & -  & -  & N12 \\
051          & 00:41:54.62 & +40:47:13.8 & N     & -  & -  & N12 \\
052          & 00:42:03.94 & +41:25:54.5 & N     & -  & -  & N12 \\
053          & 00:42:11.16 & +40:56:48.0 & N     & -  & -  & N12 \\
054          & 00:42:13.21 & +40:50:51.8 & N     & -  & -  & N12 \\
055          & 00:42:14.36 & +41:25:42.3 & N     & -  & -  & N12 \\
056          & 00:42:34.42 & +41:30:24.2 & N     & -  & -  & N12 \\
057          & 00:42:38.90 & +41:00:02.0 & N     & -  & -  & N12 \\
058          & 00:42:40.81 & +41:02:41.6 & N     & -  & -  & N12 \\
059          & 00:42:42.03 & +41:23:14.9 & Y     & S  & S  & N12 \\
060          & 00:42:42.33 & +41:39:22.7 & N     & -  & -  & N12 \\
061          & 00:42:47.12 & +40:56:57.1 & N     & -  & -  & N12 \\
062          & 00:42:49.84 & +41:02:15.7 & N     & -  & -  & N12 \\
063          & 00:42:54.65 & +41:02:26.1 & N     & -  & -  & N12 \\
064          & 00:42:56.05 & +41:35:43.7 & N     & -  & -  & N12 \\
065          & 00:42:56.85 & +41:38:37.2 & N     & -  & -  & N12 \\
066          & 00:42:57.62 & +41:37:27.8 & N     & -  & -  & N12 \\
067          & 00:43:02.05 & +41:37:46.7 & N     & -  & -  & N12 \\
068          & 00:43:02.76 & +41:37:53.6 & N     & -  & -  & N12 \\
069          & 00:43:04.34 & +41:22:23.4 & Y     & S  & S  & N12 \\
070          & 00:43:06.11 & +41:38:13.7 & N     & -  & -  & N12 \\
071          & 00:43:08.25 & +41:37:36.3 & N     & -  & -  & N12 \\
072          & 00:43:10.19 & +41:37:44.0 & N     & -  & -  & N12 \\
073          & 00:43:16.44 & +41:45:12.4 & N     & -  & -  & N12 \\
074          & 00:43:21.48 & +41:41:55.5 & N     & -  & -  & N12 \\
075          & 00:43:24.95 & +41:18:03.2 & Y     & S  & S  & N12 \\
076          & 00:43:25.68 & +41:18:30.2 & Y     & S  & S  & N12 \\
077          & 00:43:27.92 & +41:42:07.3 & N     & -  & -  & N12 \\
078          & 00:43:28.06 & +41:42:12.3 & N     & -  & -  & N12 \\
079          & 00:43:30.76 & +41:27:34.7 & Y     & S  & S  & N12 \\
080          & 00:43:31.17 & +41:12:03.5 & Y     & M  & S  & N12 \\
081          & 00:43:34.92 & +41:09:53.2 & Y     & M  & M  & N12 \\
082          & 00:43:36.51 & +41:23:15.1 & Y     & S  & S  & N12 \\
083          & 00:43:37.10 & +41:42:37.1 & N     & -  & -  & N12 \\

\bottomrule
\end{tabularx}
\end{table}
\begin{table}[H]\ContinuedFloat
\caption{\textit{Cont.}}
\begin{tabularx}{\textwidth}{CCCCCCC}
\toprule
\textbf{(1)}          & \textbf{(2)}     & \textbf{(3)}   & \textbf{(4)}         & \textbf{(5)}           & \textbf{(6)}            & \textbf{(7)} \\
\textbf{ID}           & \textbf{RA}      & \textbf{Dec}   & \textbf{F275W}       & \textbf{n(}\boldmath{$3\farcs3$}\textbf{)} & \textbf{n(}\boldmath{$1\farcs1$}\textbf{)}  & \textbf{Ref} \\
\midrule

084          & 00:43:41.72 & +41:23:04.2 & Y     & M  & S  & N12 \\
085          & 00:43:44.48 & +41:11:42.0 & Y     & M  & S  & N12 \\
086          & 00:43:47.01 & +41:12:38.0 & Y     & M  & M  & N12 \\
087          & 00:43:49.72 & +41:12:43.4 & Y     & M  & M  & N12 \\

088          & 00:43:53.34 & +41:46:38.9 & N     & -  & -  & N12 \\
089          & 00:43:57.31 & +41:48:46.2 & N     & -  & -  & N12 \\
090          & 00:44:03.39 & +41:15:18.8 & N     & -  & -  & N12 \\
091          & 00:44:06.39 & +41:19:21.0 & Y     & S  & S  & N12 \\
092          & 00:44:06.68 & +41:16:12.0 & Y     & M  & S  & N12 \\
093          & 00:44:08.58 & +41:21:21.2 & Y     & S  & S  & N12 \\
094          & 00:44:10.17 & +41:32:53.1 & Y     & M  & S  & N12 \\
095          & 00:44:10.91 & +41:16:23.2 & Y     & S  & S  & N12 \\
096          & 00:44:12.44 & +41:29:41.7 & Y     & M  & S  & N12 \\
097          & 00:44:13.06 & +41:19:20.5 & Y     & M  & S  & N12 \\
098          & 00:44:15.77 & +41:19:52.5 & Y     & S  & S  & N12 \\
099          & 00:44:16.20 & +41:21:03.5 & Y     & M  & M  & N12 \\
100          & 00:44:19.92 & +41:17:51.2 & Y     & M  & M  & N12 \\
101          & 00:44:20.58 & +41:54:12.5 & N     & -  & -  & N12 \\
102          & 00:44:22.24 & +41:18:58.4 & Y     & M  & S  & N12 \\
103          & 00:44:25.10 & +41:20:50.0 & Y     & M  & M  & N12 \\
104          & 00:44:25.43 & +41:20:44.7 & Y     & M  & M  & N12 \\
105          & 00:44:25.83 & +41:50:19.4 & N     & -  & -  & N12 \\
106          & 00:44:26.17 & +41:22:10.8 & Y     & M  & S  & N12 \\
107          & 00:44:26.32 & +41:34:19.8 & Y     & S  & S  & N12 \\
108          & 00:44:29.89 & +41:20:21.6 & Y     & M  & M  & N12 \\
109          & 00:44:30.04 & +41:52:37.1 & Y     & M  & M  & N12 \\
110          & 00:44:34.57 & +41:24:24.4 & Y     & S  & S  & N12 \\
111          & 00:44:35.15 & +41:25:45.2 & Y     & M  & S  & N12 \\
112          & 00:44:36.22 & +41:22:57.3 & Y     & S  & S  & N12 \\
113          & 00:44:36.52 & +41:22:02.0 & Y     & S  & S  & N12 \\
114          & 00:44:37.61 & +41:52:03.3 & Y     & S  & S  & N12 \\
115          & 00:44:43.10 & +41:26:19.6 & Y     & M  & M  & N12 \\
116          & 00:44:44.00 & +41:27:39.9 & Y     & M  & S  & N12 \\
117          & 00:44:44.87 & +41:28:00.7 & Y     & M  & S  & N12 \\
118          & 00:44:45.61 & +41:28:06.5 & Y     & S  & S  & N12 \\
119          & 00:44:45.90 & +41:58:03.8 & N     & -  & -  & N12 \\
120          & 00:44:49.41 & +41:30:20.7 & Y     & S  & S  & N12 \\
121          & 00:44:51.98 & +41:29:11.6 & Y     & M  & M  & N12 \\
122          & 00:44:53.06 & +41:26:01.7 & Y     & S  & S  & N12 \\
123          & 00:44:53.52 & +41:53:54.3 & Y     & M  & M  & N12 \\
124          & 00:44:55.63 & +41:31:05.1 & Y     & M  & M  & N12 \\
125          & 00:44:55.82 & +41:29:19.2 & Y     & S  & S  & N12 \\
126          & 00:45:00.96 & +41:30:58.8 & Y     & M  & M  & N12 \\
127          & 00:45:02.78 & +41:55:33.7 & Y     & M  & M  & N12 \\
128          & 00:45:06.50 & +41:34:25.1 & N     & -  & -  & N12 \\
129          & 00:45:09.18 & +41:40:21.4 & Y     & M  & S  & N12 \\
130          & 00:45:10.39 & +41:36:46.6 & Y     & M  & M  & N12 \\
131          & 00:45:11.21 & +42:05:21.7 & N     & -  & -  & N12 \\
132          & 00:45:11.27 & +41:38:15.3 & Y     & M  & S  & N12 \\
133          & 00:45:13.68 & +41:37:42.5 & Y     & M  & M  & N12 \\
134          & 00:45:14.10 & +41:37:35.2 & Y     & S  & S  & N12 \\
135          & 00:45:17.56 & +41:39:22.0 & Y     & M  & M  & N12 \\
136          & 00:45:17.89 & +41:52:09.5 & Y     & M  & S  & N12 \\
137          & 00:45:20.80 & +41:51:00.0 & Y     & S  & S  & N12 \\

\bottomrule
\end{tabularx}
\end{table}
\begin{table}[H]\ContinuedFloat
\caption{\textit{Cont.}}
\begin{tabularx}{\textwidth}{CCCCCCC}
\toprule
\textbf{(1)}          & \textbf{(2)}     & \textbf{(3)}   & \textbf{(4)}         & \textbf{(5)}           & \textbf{(6)}            & \textbf{(7)} \\
\textbf{ID}           & \textbf{RA}      & \textbf{Dec}   & \textbf{F275W}       & \textbf{n(}\boldmath{$3\farcs3$}\textbf{)} & \textbf{n(}\boldmath{$1\farcs1$}\textbf{)}  & \textbf{Ref} \\
\midrule

138          & 00:45:22.78 & +42:03:18.2 & Y     & S  & S  & N12 \\
139          & 00:45:24.26 & +41:53:52.5 & Y     & S  & S  & N12 \\
140          & 00:45:30.61 & +41:46:39.2 & Y     & S  & S  & N12 \\
141          & 00:45:31.16 & +42:06:58.1 & N     & -  & -  & N12 \\
142          & 00:45:37.05 & +41:43:02.4 & Y     & M  & S  & N12 \\
143          & 00:45:37.10 & +41:42:01.4 & Y     & M  & M  & N12 \\
144          & 00:45:37.28 & +41:43:05.2 & Y     & M  & S  & N12 \\
145          & 00:45:39.15 & +41:47:54.0 & Y     & M  & S  & N12 \\
146          & 00:45:39.35 & +41:44:39.8 & Y     & M  & S  & N12 \\
147          & 00:45:41.29 & +41:55:04.0 & Y     & M  & S  & N12 \\
148          & 00:45:42.26 & +41:45:10.1 & Y     & M  & S  & N12 \\
149          & 00:45:44.65 & +41:52:41.0 & Y     & M  & S  & N12 \\
150          & 00:45:51.12 & +42:10:15.4 & N     & -  & -  & N12 \\
151          & 00:45:51.35 & +41:42:42.0 & N     & -  & -  & N12 \\
152          & 00:46:28.51 & +42:11:27.6 & Y     & S  & S  & N12 \\
153          & 00:46:45.02 & +42:13:01.3 & Y     & S  & S  & N12 \\
154          & 00:46:55.89 & +42:00:15.1 & Y     & M  & M  & N12 \\
155          & 00:40:19.66 & +40:42:32.5 & N     & -  & -  & N23 \\
156          & 00:40:31.21 & +40:41:28.1 & N     & -  & -  & N23 \\
157          & 00:40:32.57 & +40:39:01.3 & N     & -  & -  & N23 \\
158          & 00:40:39.59 & +40:44:49.5 & N     & -  & -  & N23 \\
159          & 00:40:42.44 & +40:45:05.3 & N     & -  & -  & N23 \\
160          & 00:40:54.05 & +40:37:08.3 & N     & -  & -  & N23 \\
161          & 00:43:26.06 & +41:42:60.0 & N     & -  & -  & N23 \\
162          & 00:43:38.95 & +41:43:27.0 & N     & -  & -  & N23 \\
163          & 00:43:41.15 & +41:44:13.6 & N     & -  & -  & N23 \\
164          & 00:43:53.03 & +41:21:41.0 & Y     & S  & S  & N23 \\
165          & 00:43:59.44 & +41:48:23.9 & N     & -  & -  & N23 \\
166          & 00:44:06.41 & +41:20:20.8 & Y     & S  & S  & N23 \\
167          & 00:44:08.13 & +41:21:00.6 & Y     & M  & S  & N23 \\
168          & 00:44:13.56 & +41:20:04.7 & Y     & M  & M  & N23 \\
169          & 00:44:18.10 & +41:18:50.8 & Y     & M  & M  & N23 \\
170          & 00:44:25.09 & +41:20:46.4 & Y     & M  & M  & N23 \\
171          & 00:44:31.39 & +41:21:14.0 & Y     & M  & S  & N23 \\
172          & 00:44:33.91 & +41:25:01.2 & Y     & M  & S  & N23 \\
173          & 00:44:38.04 & +41:25:18.7 & Y     & M  & S  & N23 \\
174          & 00:43:43.61 & +41:45:27.9 & N     & -  & -  & S16 \\
\bottomrule
\end{tabularx}
\end{table}
\vspace{-9pt}

\begin{figure}[H]

\includegraphics[width=0.9\textwidth]{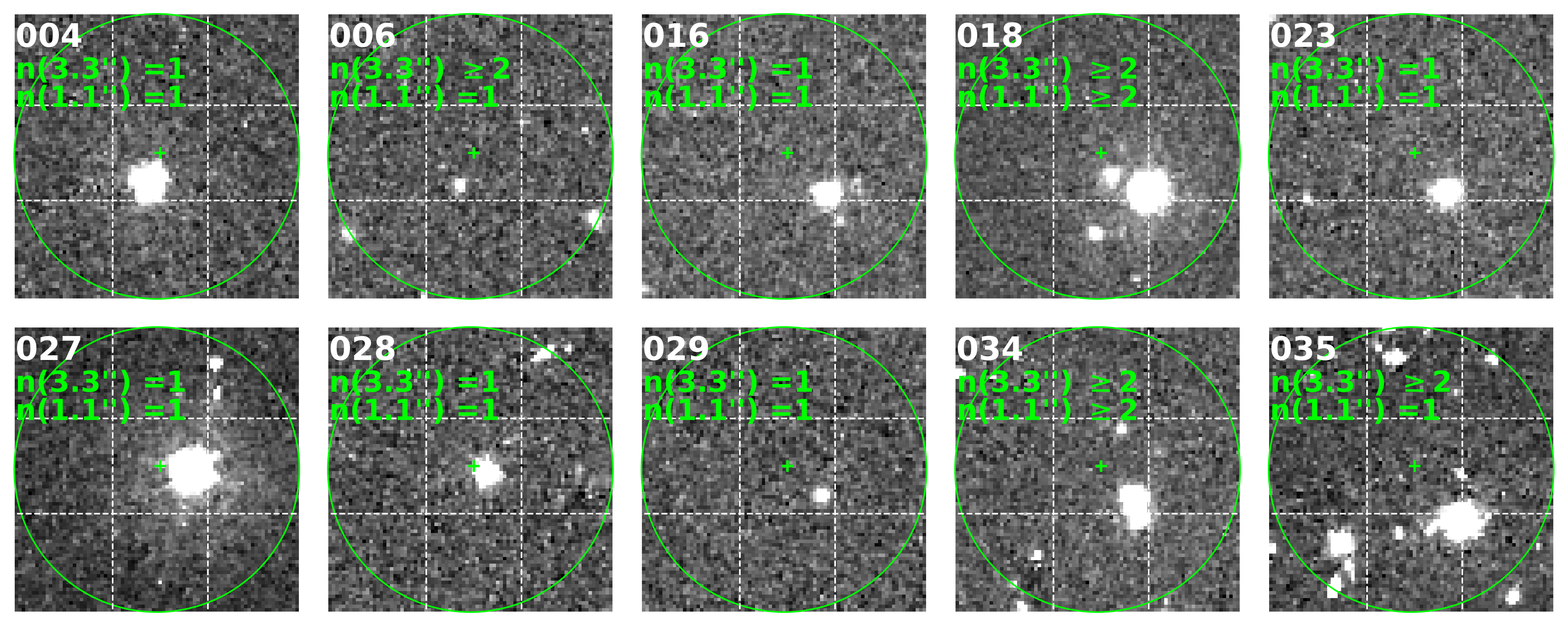}
\caption{\textit{Cont.}}
\end{figure}
\begin{figure}[H]\ContinuedFloat
\includegraphics[width=0.9\textwidth]{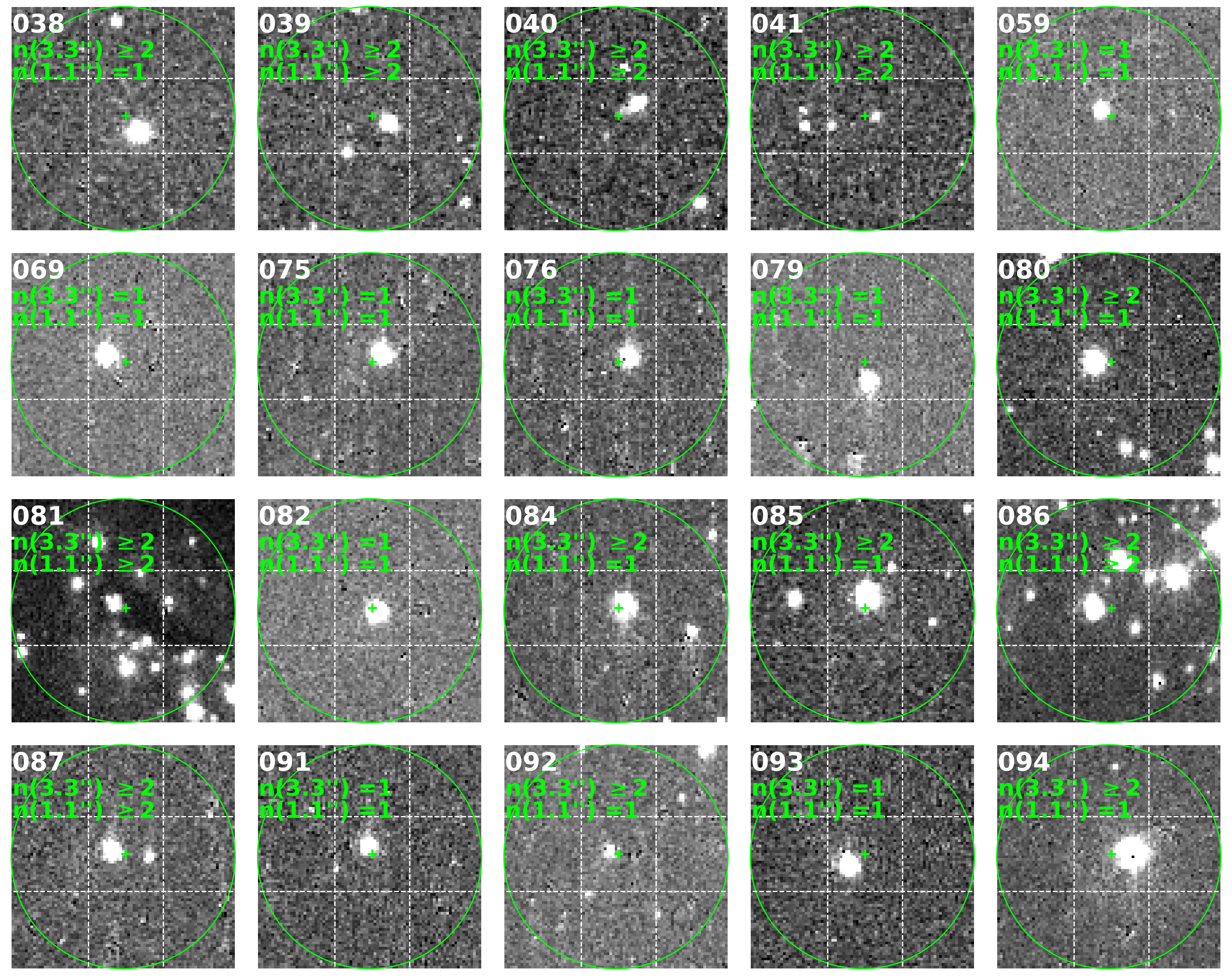}
\includegraphics[width=0.9\textwidth]{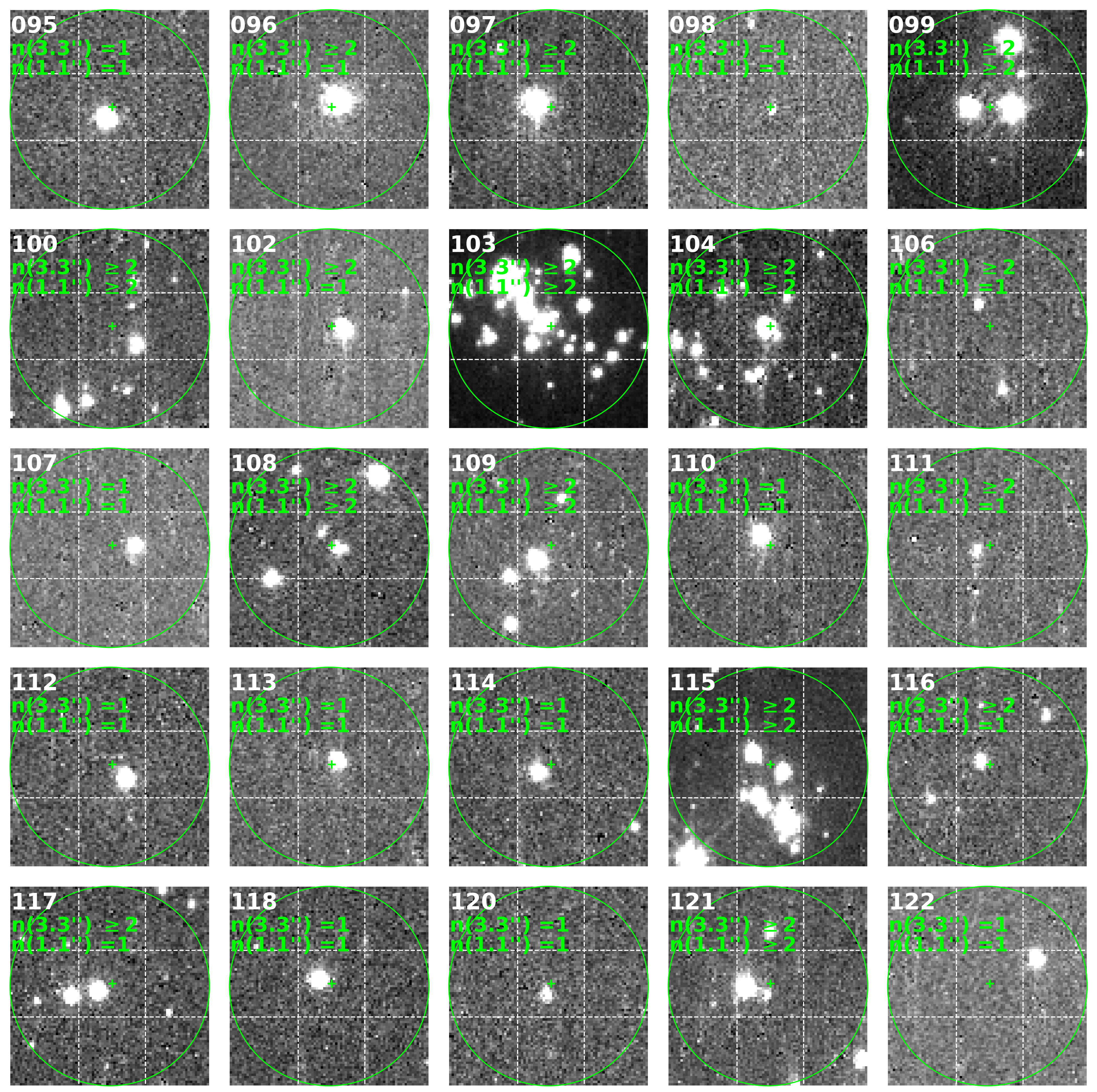}
\caption{\textit{Cont.}}
\end{figure}
\begin{figure}[H]\ContinuedFloat
\includegraphics[width=0.9\textwidth]{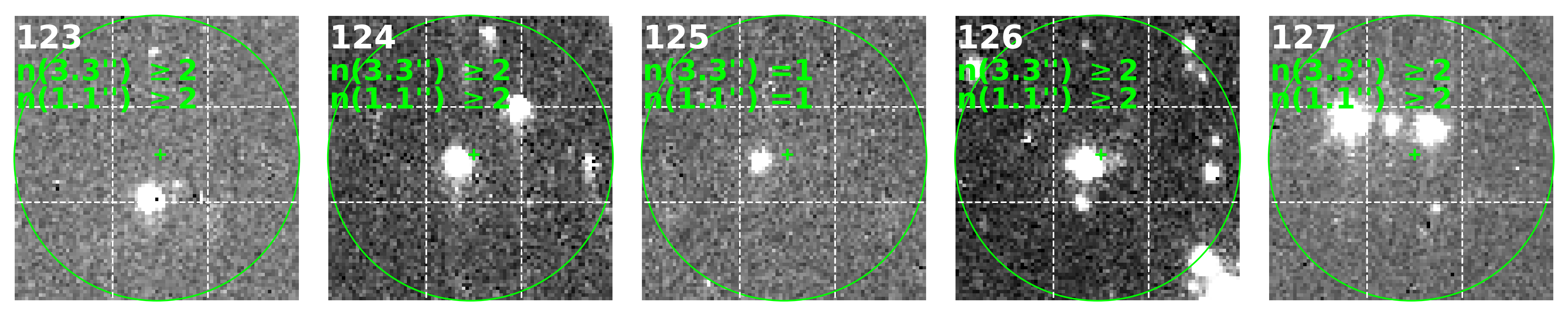}
\includegraphics[width=0.9\textwidth]{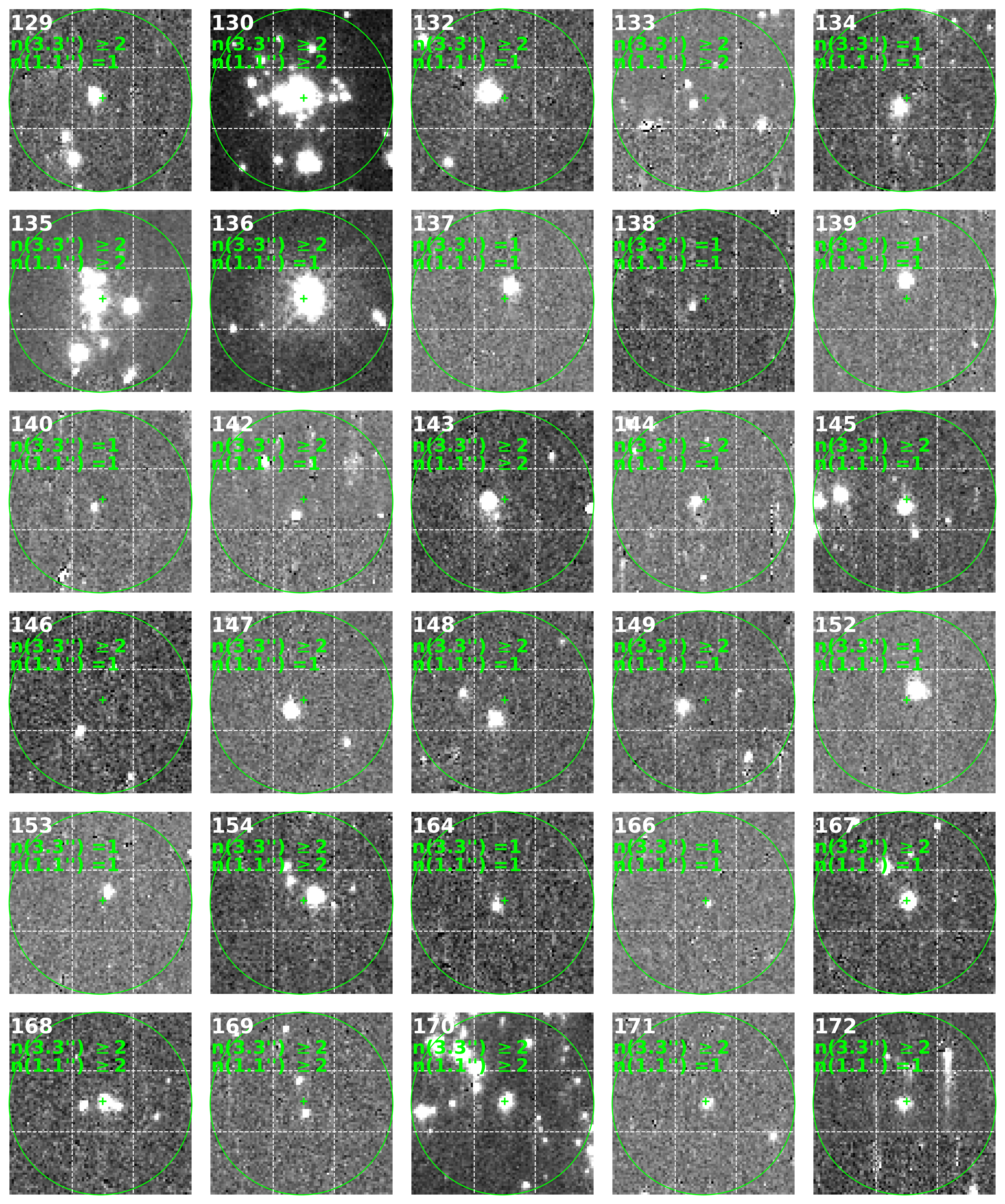}
\includegraphics[width=0.9\textwidth]{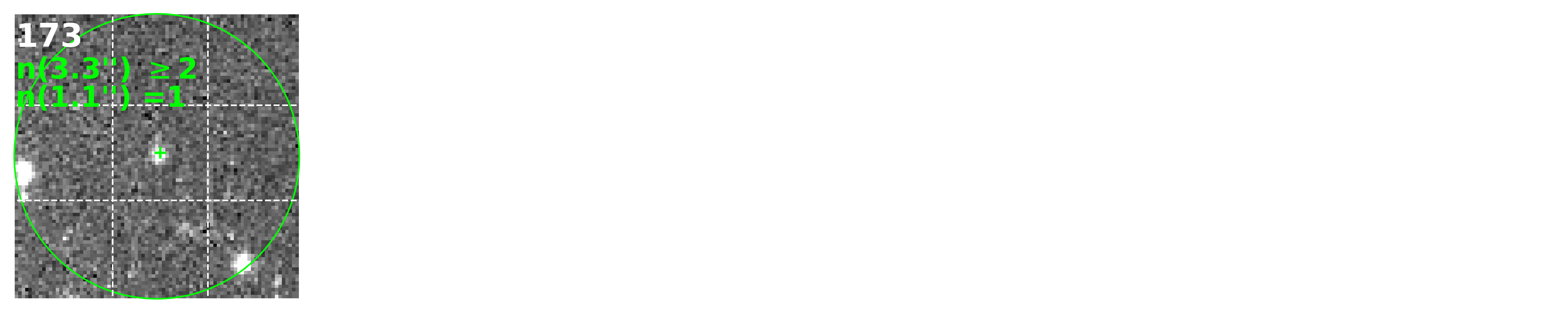}
\caption{HST F275W images for matched targets. In~each panel, ID is shown in the upper left. The~green circle, with~a diameter of $3\farcs3$, represents the position of the LAMOST fiber. The~green cross marks the target center coordinates, of~which (RA, Dec) are listed in the Table~\ref{tbl:obs_infor}. The~white dashed lines divide the image into 3 $\times$ 3 grids, with~each grid covering $1\farcs1$. N($3\farcs3$) and n($1\farcs1$) denote the number of bright sources in the FOV of $3\farcs3$ and $1\farcs1$, respectively.}
\label{fig:merged_hstimg1}

\end{figure}

\clearpage 
\begin{adjustwidth}{-\extralength}{0cm}

\reftitle{References}

\PublishersNote{}
\end{adjustwidth}
\end{document}